\documentclass[11pt]{article}
\usepackage{amssymb,amsmath,amsfonts, mathtools}
\usepackage{enumerate}
\usepackage{frcursive}
\usepackage[T1]{fontenc}

\newenvironment{frcseries}{\fontsize{9pt}{9pt}\fontfamily{frc}\selectfont}{}
\newcommand{\textfrc}[1]{{\frcseries#1}}
\newcommand{\h}{\text{\textfrc{h}}}
\newenvironment{frcseriesa}{\fontsize{7pt}{7pt}\fontfamily{frc}\selectfont}{}
\newcommand{\textfrca}[1]{{\frcseriesa#1}}
\newcommand{\ve}{\varepsilon}
\newcommand{\hh}{\text{\textfrca{h}}}
\usepackage[sort,compress]{cite}
\usepackage{amsfonts,euscript,amssymb,stmaryrd,braket}
\usepackage{graphics,tikz}
\usetikzlibrary{arrows,decorations.markings,patterns}
\usepackage{slashed}
\definecolor{hyperref}{RGB}{026,028,185}
\usepackage[bookmarks=true,colorlinks=true,linkcolor=hyperref,citecolor=hyperref,urlcolor=hyperref,bookmarksnumbered]{hyperref}
\usepackage{cite}

\def\clock{{\count0=\time
           \divide\count0 60
           \ifnum\count0<10 0\fi\the\count0
           \multiply\count0 -60 \advance\count0 \time
           :\ifnum\count0<10 0\fi \the\count0
         }}
\newcommand{\timestamp}{{\small\vbox{\hbox{\tt\jobname.tex}
\hbox{\the\day/\the\month/\the\year, \clock}}}}

\newcommand{\g}{\gamma}

\newcommand{\C}{\mathbb{C}}

\newcommand{\nn}{\nonumber}
\newcommand{\ba}{\begin{eqnarray}}
\newcommand{\ea}{\end{eqnarray}}

\DeclareMathOperator{\diag}{diag}

\newcommand{\ads}{\textup{\textrm{AdS}}}

\newcommand{\sphere}{\textup{\textrm{S}}}

\newcommand{\be}{\begin{equation}}
\newcommand{\ee}{\end{equation}}

\newcommand{\Section}[1]{\section{#1}\setcounter{equation}{0}}

\makeatletter
\let\old@startsection=\@startsection
\let\oldl@section=\l@section
\renewcommand{\@startsection}[6]{\old@startsection{#1}{#2}{#3}{#4}{#5}{#6\mathversion{bold}}}
\renewcommand{\l@section}[2]{\oldl@section{\mathversion{bold}#1}{#2}}
\makeatother

\numberwithin{equation}{section}

\newcommand{\grp}[1]{\mathrm{#1}}

\newcommand{\grSO}{\grp{SO}}

\newcommand{\quarter}{\frac{1}{4}}
\newcommand{\half}{\frac{1}{2}}
\newcommand{\p}{\partial}

\usepackage{color}

\def\s{\sigma}

\begin{document}
\renewcommand{\thefootnote}{\arabic{footnote}}

\overfullrule=0pt
\parskip=2pt
\parindent=12pt
\headheight=0in \headsep=0in \topmargin=0in \oddsidemargin=0in

\vspace{ -3cm} \thispagestyle{empty} \vspace{-1cm}
\begin{flushright} 
\footnotesize
HU-EP-15/25\\
\end{flushright}%

\begin{center}
\vspace{1.2cm}
{\Large\bf \mathversion{bold}
Remarks on the geometrical properties  \\ 
of semiclassically  quantized  strings 
}

 \vspace{0.8cm} {
  V.~Forini$^{a,}$\footnote{ {\tt $\{$valentina.forini,edoardo.vescovi$\}$@\,physik.hu-berlin.de}},
 V.~Giangreco~M. Puletti$^{b,}$\footnote{ {\tt vgmp@hi.is}},
  L.~Griguolo$^{c,}$\footnote{ {\tt luca.griguolo@fis.unipr.it}}, 
D.~Seminara$^{d,}$\footnote{ {\tt seminara@fi.infn.it}},
E.~Vescovi$^{a,e,1}$}
 \vskip  0.5cm

\small
{\em
$^{a}$Institut f\"ur Physik, Humboldt-Universit\"at zu Berlin, IRIS Adlershof, \\Zum Gro\ss en Windkanal 6, 12489 Berlin, Germany  
\vskip 0.05cm
$^{b}$ University of Iceland,
Science Institute,
Dunhaga 3,  107 Reykjavik, Iceland
  \vskip 0.05cm
$^{c}$ Dipartimento di Fisica e Scienze della Terra, Universit\'a di Parma and INFN
Gruppo Collegato di Parma, Viale G.P. Usberti 7/A, 43100 Parma, Italy
  \vskip 0.05cm
$^{d}$ Dipartimento di Fisica, Universit\'a di Firenze and INFN Sezione di Firenze, Via G. Sansone 1, 50019 Sesto Fiorentino,
Italy
\vskip 0.05cm
$^{e}$Perimeter Institute for Theoretical Physics, Waterloo, Ontario N2L 2Y5, Canada
}
\normalsize


\end{center}

\vspace{0.3cm}
\begin{abstract}

We discuss some geometrical aspects of the semiclassical quantization of string solutions in Type IIB Green-Schwarz action on $\ads_5\times \sphere^5$. We concentrate on quadratic fluctuations around classical  configurations, expressing the relevant differential operators in terms of  (intrinsic and extrinsic) invariants of the background geometry.  The aim of our exercises is to present some compact expressions encoding the spectral properties of bosonic and fermionic fluctuations. The appearing of non-trivial structures on the relevant bundles and their role in concrete computations are also considered. We corroborate the presentation of general formulas by working out explicitly a couple of relevant examples, namely the spinning string and the latitude BPS Wilson loop.

\end{abstract}

\newpage
\tableofcontents
\setcounter{footnote}{0}
  \section{Introduction}
 
The geometric properties of string worldsheets embedded in a $D$-dimensional space-time, and of linearized perturbations above them, have been object of various studies since the seminal observation on the relevance of quantizing string models~\cite{Polyakov:1981rd}.
In the framework of the AdS/CFT correspondence, the semiclassical study of strings in non-trivial backgrounds~\cite{Gubser:2002tv} has played a crucial role, expecially in connection with the detection of the underlying integrable structure~\cite{Bena:2003wd}.  At the same time it provides a powerful tool to check, at strong coupling,  exact QFT results obtained through localization procedure \cite{Pestun:2007rz,Kapustin,Marino:2009jd,Drukker:2010nc}  for BPS 
\cite{Erickson:2000af,Drukker:2000ep,Zarembo:2002an,Drukker:2007qr,Buchel:2013id,Bigazzi:2013xia} and non-BPS observables~
\cite{Tseytlin:2010jv,McLoughlin:2010jw,Alday:2010kn,Klose:2010ki,Drukker:2011za,Correa:2012at,Forini:2012bb,Buchbinder:2013nta, Aguilera-Damia:2014bqa}~\footnote{The precise match between results obtained via semiclassical quantization and the exact prediction obtained via supersymmetric localization does not go beyond leading order in $\sigma$-model perturbation theory, see~\cite{Kruczenski:2008zk,Kristjansen:2012nz}.}. 
As a matter of fact a large variety of classical string solutions have been proposed to correspond to CFT gauge-invariant operators, Wilson loops or dimensionally reduced amplitudes, the original suggestions being supported beyond the leading classical order by sometimes non-trivial calculations at one-loop order~\cite{Forste:1999qn,Drukker:2000ep,
Frolov:2002av, Frolov:2003qc,Frolov:2003tu,Frolov:2004bh,Park:2005ji,Beisert:2005mq,Kruczenski:2008zk,Beccaria:2008tg,Beccaria:2010ry,Forini:2010ek,Drukker:2011za,Forini:2012bb,Forini:2014kza,Aguilera-Damia:2014bqa} (see also~\cite{Giombi:2010fa, Bianchi:2014ada} for higher order computations in special cases, the so-called homogenous  solutions, for which derivatives of the background fields are constant).

\bigskip

The natural setting in which these analyses have been performed is the Green-Schwarz $\sigma$-model on AdS$_5\times$S$^5$ \cite{Metsaev:1998it}, the relevant string background for ${\cal N}=4$ Super Yang-Mills gauge theory. The first step in order to compute one-loop quantum corrections to classical solutions of the string $\sigma$-model is of course to derive the quadratic action for the small fluctuations. Then, after appropriate gauge-fixings conveniently chosen according to the original form of the action (Polyakov or Nambu-Goto) and a careful definition of the path-integral measure, the problem is reduced to the evaluation of a bunch of bosonic and fermionic functional determinants. The geometry of the classical string background is encoded into the structure of the differential operators entering the computations and in the possible appearance of zero-modes, affecting the integration measure. Finally a regularization procedure, compatible with the symmetry of the specific problem, should be exploited to derive sensible results from the formal expression of the one-loop effective action. This  project was first addressed in~\cite{Drukker:2000ep}, where a systematic treatment of the Green-Schwarz (GS) string in curved AdS$_5\times$S$^5$ space was initiated and the quadratic fluctuation operators in conformal and static gauges (for Polyakov and Nambu-Goto actions) were found. A careful treatment of the measure factors and ghost determinants was also presented. The aim of that analysis was mainly focussed on the study of particular minimal surfaces, namely  the ones associated to the straight, the circular and the antiparallel lines Wilson loops. Some relevant formulas were somehow tailored on the specific examples and on the use of the conformal gauge for the Polyakov path-integral. This pioneering paper was followed along the years by many investigations, both on the open string side (mainly minimal surfaces associated to BPS and non-BPS Wilson loops) and on the closed string side (for example different classes of string solutions related to CFT operators). Although a large number of results were obtained from one-loop quantum corrections, sometimes brilliantly confirming the expectations from integrability and localization, all this analysis relied somehow on the particular form of the classical string configuration. To further increase the effective power of this approach, it would be desirable to have a general and manifestly \emph{covariant} formalism to describe fluctuations, which should be also independent of the particular string solution or of the background in which the string is embedded. Here we attempt some modest steps in this direction, collecting and generalizing some useful formulas previously appeared in the literature and presenting the results of a series of exercises that we hope interesting for people working in the field.    

The central point of our analysis is the application of some elementary concepts of intrinsic and extrinsic geometry to the properties of string worldsheet embedded in a $D$-dimensional curved space-time. We take full advantage of the equations of Gauss, Codazzi, and Ricci for  surfaces embedded in a general background to obtain simple and general expressions for perturbations over them.  We follow and enlarge earlier investigations~\cite{Callan:1989nz,Drukker:2000ep}~\footnote{See also~\cite{Capovilla:1994bs,Viswanathan:1996yg} and references therein, where this analysis has been exploited for the description of QCD strings or stability effects for membrane solutions, and the more recent~\cite{Faraggi:2011ge}.}, starting from the Polyakov formulation and trying to present a systematic and self-consistent perspective to the study of fluctuations in the AdS$_5\times$S$^5$. The main result consists in general formulas for bosonic and fermionic fluctuation operators above a classical string solution: expressions as \eqref{Mtot}-\eqref{ms5mads5}-\eqref{sumrulebos} and \eqref{lagrfermkin}-\eqref{L_ferm_flux}-\eqref{gammastartilde_better}-\eqref{trm_fermions}-\eqref{sum_rule_all} only require as an input generic properties of the classical configuration and basic information about the space-time background. The inclusion of a suitable choice of orthonormal vectors which are orthogonal to the surface spanned by the string solution will also play a major role. In particular, after fixing the conformal gauge, it allows to decouple explicitly the longitudinal modes arriving to a final expression, similar to ones that would be obtained in the Nambu-Goto formalism in the static gauge for bosons. In the fermionic sector the reduction of Green-Schwarz fermions to a set of two-dimensional Dirac spinor is equivalently accomplished. 
We  provide explicit expressions in terms of geometric invariants for bosonic and fermionic ``masses'', noticing that in all the cases previously analyzed simplifications occur which are associated to the flatness of the normal bundle. While in the bosonic case similar formulas appeared before scattered in the literature, our treatment of the fermionic case, due to the complications related to the flux term, is somehow novel in its generality. 

To proceed in the one-loop analysis, one has then to compute the functional determinants associated to the fluctuations operators, which can be done with standard methods for functional determinants  (see for example~\cite{Kruczenski:2008zk,Beccaria:2010ry,Forini:2010ek,Drukker:2011za,Forini:2014kza}) and could involve many regularization subtleties~\cite{FGGSVlatitude}. Here we do not address the problem of regularization procedure and other important issues, as the appropriate definition of integration measure, $\kappa$-symmetry ghosts, Jacobians due to change of fluctuation basis are also left to future investigations. These topics should deserve a careful study, expecially when BPS configurations are considered and the quantum fluctuations must preserve this property. 

A natural generalization of our investigations concerns Type IIA and IIB string backgrounds relevant for the AdS$_4$/CFT$_3$ and AdS$_3$/CFT$_2$ correspondences respectively: we expect it should be possible simply exploiting some general features of their geometry. For example, backgrounds like  $\ads_4\times \C P^3$,  $\ads_3\times \sphere^3\times M^4$, $\ads_3\times \sphere^2\times M^5$, $\ads_2\times \sphere^2\times M^6$ (where $M^4=T^4, \sphere^3\times \sphere^1$, $M^5=\sphere^3\times T^2$) are direct products of symmetric spaces, which results in a structure of the Riemann tensor resembling the ``separability'' of \eqref{riemannmaxsusy} and  allowing the writing of formulas similar to \eqref{massesads5s5}-\eqref{S5-V-AdS5}.

In the perspective adopted in this paper, there is also no explicit reference to the classical integrability of the Green-Schwarz superstring on $\ads_5\times \sphere^5$~\cite{Bena:2003wd}. In a number of semiclassical studies~\cite{Beccaria:2010ry,Drukker:2011za,Forini:2014kza} the underlying integrable structure of the $\ads_5\times \sphere^5$ background emerges, for example, in the appearance of certain special, integrable, differential operators~\cite{Beccaria:2010ry, Forini:2014kza}, whose  determinants can be calculated explicitly and result in closed (albeit in integral form) expressions for the one loop partition functions~\footnote{In another kind of perturbative analysis of the worldsheet $\sigma$-model, i.e. the perturbative evaluation of the massive S-matrix for the elementary excitations around the BMN vacuum~\cite{Klose:2006zd, Klose:2007rz,Puletti:2007hq} (see also~\cite{Arutyunov:2009ga,McLoughlin:2010jw} for reviews), the one-loop computation for the full $AdS_5\times S^5$ case~\cite{Bianchi:2013nra,Engelund:2013fja,Roiban:2014cia} (in a certain regularization scheme) reproduces exactly the results~\cite{Beisert:2005tm} predicted by (symmetries and) quantum integrability.}.  The question of  a deeper relation between such geometric approach to fluctuations and the integrability of the $\sigma$-model of interest should become more manifest within the algebraic curve approach to semiclassical quantization~\cite{Beisert:2005bv,Gromov:2009zza,Vicedo:2008jy}, likely on the lines of~\cite{Ishizeki:2011bf} and is an interesting issue to be addressed in the future.

\bigskip 
 
The paper proceeds as follows. The geometrical formulation of classical string solutions as minimal surfaces is briefly recalled in Section \ref{sec:classicalstring}. In Section \ref{sec:Bosonic Fluctuations} we discuss the bosonic sector. After reviewing previous analysis based on background field method for nonlinear $\sigma$-models and  the expansion in  normal coordinates, we write the relevant contributions in terms of intrinsic and extrinsic geometric invariants of the classical solution. We discuss the gauge-fixing and the decoupling of the longitudinal modes, as well as the arising of gauge connections in the covariant derivatives associated to the structure of normal bundle. The spectral properties are also investigated, obtaining the mass matrix and deriving some sum rules. Section \ref{sec:Fermionic fluctuations} is instead devoted to the fermionic sector. We explicitly obtain the fermionic kinetic terms by performing suitable rotations that reduce the GS spinor to two-dimensional Dirac fermions and observe the arising of normal bundle gauge connection as in the bosonic case. Then we discuss the mass matrix that, after a careful treatment of the flux contribution, is expressed in terms of geometrical invariants. We conclude with Section \ref{sec:applications}, where a couple of relevant situations (the well-known spinning string solution of~\cite{Gubser:2002tv,Frolov:2002av} and fluctuations over the minimal surface associated to the 1/4 BPS Wilson loop operator~\cite{Drukker:2005cu, Drukker:2006ga, Drukker:2007qr}) are considered,  in which the general structures previously derived are exemplified.  


 \Section{The background equation}
\label{sec:classicalstring}

We start by recalling some basic fact about classical string theory. In particular we will review the statement that classical string 
solutions are minimal surfaces, \emph{i.e.} surfaces of vanishing mean curvature. 
We shall deal with  classical backgrounds which are extrema of the Nambu-Goto action (fermions are of course zero at classical level)
\be
S_{N.G.}=\int_{\Sigma} d^{2}\sigma \sqrt{\gamma},
\ee
where  $\gamma_{\alpha\beta}\equiv G_{mn}\partial_{\alpha} X^{m}\partial_{\beta}X^{n}$ $(\alpha,\beta=1,2)$ is the induced metric, namely the pull-back of  the target space metric $G_{mn}$ $(l\,,m\,,n\,,p \,,.\,. \,. \,=1, \dots\, ,D)$ on the worldsheet $\Sigma$.  The   background $X^{m}$ solves the  Euler-Lagrangian equation 
\be
\begin{split}
\partial_{\alpha}\left(\sqrt{\gamma}\gamma^{\alpha\beta} G_{mn}\partial_{\beta}X^{n}\right) -
\frac{1}{2} \sqrt{\gamma}\gamma^{\alpha\beta} \partial_{m} G_{np}\partial_{\alpha} X^{n}\partial_{\beta}X^{p}=0\,,
\end{split}
\ee
which is conveniently rewritten  as follows
\be
G_{mn}\square X^{n}=\frac{1}{2} \gamma^{\alpha\beta} \partial_{m} G_{np}\partial_{\alpha} X^{n}\partial_{\beta} X^{p}-\gamma^{\alpha\beta} \partial_{\alpha} X^{p}\partial_{p} G_{m n}\partial_{\beta}
X^{n}
\ee
where $\square =\frac{1}{\sqrt{\gamma}}\,\partial_\alpha(\sqrt{\gamma}\,\gamma^{\alpha\beta}\,\partial_\beta)$ is the covariant  Laplacian on worldsheet scalars. Introducing $\Gamma^m_{np}$, the Christoffel connections for $G_{np}$, we have 
\be
\square X^{m}+\gamma^{\alpha\beta}\,\Gamma^{m}_{np} \,\partial_{\alpha} X^{n}\partial_{\beta} X^{p}=0\,.
\ee
The covariant Laplacian can be further expanded in terms of the induced metric $\gamma_{\alpha\beta}$ and the related Christoffel connections $\Lambda^{\rho}_{\alpha\beta}$ to find
\be\label{extrinsic}
\gamma^{\alpha\beta}(\partial_{\alpha}\partial_{\beta} X^{m}-\Lambda^{\rho}_{\alpha\beta}\partial_{\rho} X^{m}+\Gamma^{m}_{np}\partial_{\alpha} X^{n}\partial_{\beta} X^{p})\equiv\gamma^{\alpha\beta}
K^{m}_{\alpha\beta}=0\,,
\ee
where the  second fundamental form $K^{m}_{\alpha\beta}$ of the embedding (or {\it extrinsic curvature}) has been introduced. Then, the string equation 
of motion simply  states  that  the  mean curvature $ K^{m}$ vanishes 
\be\label{meanvanish}
 K^{m}\equiv\gamma^{\alpha\beta} K^{m}_{\alpha\beta}=0~.
\ee
As a matter of fact the extrinsic curvature is automatically orthogonal to the two  vectors $t^m_\alpha\equiv \partial_\alpha X^m$ $(\alpha=1,2)$, tangent to the worldsheet,  
\be
G_{mn} t^m_\alpha K^n_{\rho\sigma}=0.
\ee

\noindent
Physically  this means that only $D-2$ of the $D$ equations in \eqref{meanvanish} are independent and they govern the $D-2$ transverse degrees of freedom. The longitudinal ones are obviously gauge degrees of freedom.
 
 \noindent
The same result can be of course recovered from the Polyakov action
\be
S_{P.}=\int_{\Sigma} d^{2}\sigma \sqrt{h} h^{\alpha\beta} G_{mn}\partial_{\alpha} X^{m}\partial_{\beta}
X^{n}\,
\ee
 where  the $2$d worldsheet  metric $h_{\alpha\beta}$ is now an independent field.  In this case the 
 dynamical equations for the embedding coordinates $X^{m}$ are slightly different and read
\be\label{eompolyakov}
\square_{h} X^{m}+h^{\alpha\beta}\Gamma^{m}_{np}\partial_{\alpha} X^{n}\partial_{\beta} X^{p}=h^{\alpha\beta} K^{m}_{\alpha\beta}+h^{\alpha\beta} (\Lambda^{\rho}_{
\alpha\beta}-\tilde \Gamma^{\rho}_{
\alpha\beta} ) \partial_{\rho}X^{m}=0,
\ee
where $\square_{h}$ denotes the covariant Laplacian and $\tilde \Gamma$ the Christoffel symbols for the auxiliary metric $ h_{\alpha\beta}$, whereas  $\Lambda$ 
are the ones for the induced metric. But if we use that the algebraic equation for the metric field  
$h_{\alpha\beta}$  is solved by   $h_{\alpha\beta}=e^{\varphi} \gamma_{\alpha\beta}$, the last term in \eqref{eompolyakov}  vanishes,
\be
h^{\alpha\beta}(\Lambda^{\rho}_{\alpha\beta}-\tilde \Gamma^{\rho}_{
\alpha\beta} ) 
=0,
\ee
and  the string equation of motion again reduces to  \eqref{meanvanish}.

\section{Bosonic Fluctuations} 
\label{sec:Bosonic Fluctuations}

In this section we shall discuss the action for bosonic fluctuations around a classical  background. 
After reviewing previous analysis~\cite{Callan:1989nz,Drukker:2000ep} based on background field method 
for nonlinear  $\sigma$-models and the virtues of the expansion in ``geodesic'' normal coordinates
~\cite{AlvarezGaume:1981hn}, we write the relevant contributions in terms of intrinsic and extrinsic geometric invariants
of the classical solution. 

\subsection{The bosonic Lagangian}
 \label{Bosonicfluct}

We will discuss the bosonic fluctuations starting from  the Polyakov action
\be\label{polyakov2}
S=\int d^{2}\sigma\sqrt{h} h^{\alpha\beta} \partial_{\alpha} \tilde{X}^{n}\partial_{\beta} \tilde{X}^{m} G_{mn}(\tilde{X}).
\ee
A well-known subtlety of the expansion  of a non-linear $\sigma$-model around a classical background 
$X^m$~\cite{AlvarezGaume:1981hn} is that writing it as a power series in terms of fluctuations, defined as $\delta X^{m}=\tilde X^{m}-X^{n}$, 
does not lead to a manifestly covariant expression for the series coefficients. As a matter of fact the difference between 
coordinates values at nearby points of the manifold does not transform simply under reparametrization.  
The easiest way to obtain a manifestly covariant form for the coefficients is to take advantage of the method 
of normal (or Riemann) coordinates,  expressing $\delta X^m$ as a local power series in spacetime vectors, 
those tangent to the spacetime geodesic connecting $X^{m}$ with $X^{m}+\delta X^{m}$~\cite{AlvarezGaume:1981hn}.
More precisely one considers a geodesic $X^{m}(t)$  with $t$ parametrizing the arc length
such that 
\be
X^{m}(0)= X^{m} \ \ \ \ \ \mathrm{and}\ \ \ \ \   X^{m}(1)=\tilde  X^{m}.
\ee
Solving then the geodesic equation for $X^{m}(t)$
\be
\ddot{X}^{m}(t)+\Gamma^{m}_{np}\dot{X}^{n}(t) \dot{X}^{p}(t)=0
\ee
in terms of the tangent vector to this geodesic in $t=0$
\be
\zeta^{m}\equiv\dot{X}^{m}(0)
\ee
 one finds
\be
X^{m}(t)=X^{m}+t\zeta^{m}-\frac{1}{2} t^{2}\Gamma^{m}_{np}\zeta^{n}\zeta^{p}+O(t^{3}).
\ee
For $t=1$, this means~\footnote{
At quadratic level for fluctuations, the term linear in $\zeta$ does not play a crucial role. In fact it yields only
contributions which are proportional to the equation of motions:
$$
S=S_{0}+\int d^{2}\sigma \left.\frac{\delta S}{\delta \tilde X^{m}}\right|_{\tilde X=X} \zeta^{m}-
\frac{1}{2}\int d^{2}\sigma \left(\left.\frac{\delta S}{\delta \tilde X^{m}\delta \tilde X^{n}}\right|_{\tilde X=X} 
\zeta^{m}\zeta^{n}-\left.\frac{\delta S}{\delta \tilde X^{m}}\right|_{\tilde X=X} \Gamma^{m}_{ab}\zeta^{a}
\zeta^{b}\right).
$$
However, its introduction allows to simplify the algebra involved in the computation.}
\be
\tilde X^{m}=X^{m}+\zeta^{m}-\frac{1}{2} \Gamma^{m}_{np}\zeta^{n}\zeta^{p}+O(\zeta^{3})\ \ \Rightarrow\ \ 
\tilde X^{m}-X^{m}=\zeta^{m}-\frac{1}{2} \Gamma^{m}_{np}\zeta^{n}\zeta^{p}+O(\zeta^{3})\,.
\ee
 where $\Gamma^m_{np}\equiv\Gamma^m_{np}(X^m)$. The difference $\delta X^m=\tilde X^{m}-X^{m}$ 
is now the desired local power series in the vector
$\zeta^{m}$, which can then be conveniently used as a fundamental variable. 
Combining the expansions of the derivatives of the embedding coordinates
\[
\!\!\!\!
\partial_{\alpha}\tilde X^{m}\!=\! \partial_{\alpha} X^{m}\!+\!\nabla_{\alpha} \zeta^{m}\!-\!\partial_\alpha 
X^n\Gamma^{m}_{n r}\zeta^{r}\!-\!\frac{1}{2}\partial_{\alpha} X^{r}
(\partial_{r}\Gamma^{m}_{np}\!-\!2\Gamma^{m}_{nl}\Gamma^{l}_{r p})\zeta^{n}\zeta^{p}
\!-\!\Gamma^{m}_{np}\zeta^{n}\nabla_{\alpha}\zeta^{p}\!+\!O(\zeta^{3})\,,
\]
where $\nabla_\alpha\zeta^m\equiv\partial_\alpha\zeta^m+\Gamma^m_{np}\,\partial_\alpha 
X^n\,\zeta^p$, with the contribution of the target metric
\be
G_{mn}(\tilde X)=G_{mn}(X)+\left(\zeta^{r}-\frac{1}{2} \Gamma^{r}_{pq}
\zeta^{p}\zeta^{q}\right)\partial_{r} G_{mn}(X)+
\frac{1}{2}\zeta^{r}\zeta^{s}\partial_{r}\partial_{s} G_{mn}(X)+ O(\zeta^{3}),
\ee
we find 
the fluctation  action in the Polyakov formulation \eqref{polyakov2} (see~\cite{Callan:1989nz} for example)
\be
S=S^{(0)}_B(X)+\int d^{2}\sigma\sqrt{h} h^{\alpha\beta} [\nabla_{\alpha}\zeta^{m}\nabla_{\beta}\zeta^{n} G_{mn}-R_{rm,sn}\zeta^{r}\zeta^{s}\partial_{\alpha} X^{m}\partial_{\beta} X^{n}]+O(\zeta^{3}).
\ee
The term $S^{(0)}_B(X)$ denotes the classical action, while the second  one describes the quadratic fluctuations and it will be denoted with
$S^{(2)}_B$ in the following. 
In order to have a canonically normalized kinetic term it is convenient to introduce a set  of vielbein $E^{A}_{m}$  ($A,B,...=1,...,D$) for the target metric
\be
G_{mn}=\eta_{AB} E^{A}_{m} E^{B}_{n}
\ee
and a set of zweibein  $e^{a}_{\alpha} $ ($a,b,...=1,2$) for  $h_{\alpha\beta}$.  In terms of the redefined  fluctuations 
fields
\be
\xi^{A}=E^{A}_{m}\zeta^{m}\,,
\ee
the quadratic fluctuation action for bosons becomes~\cite{Callan:1989nz,Drukker:2000ep}
\be\label{actionDGT}
S_{2B}=\int d^{2}\sigma\sqrt{h}  \,[\,h^{\alpha\beta} D_{\alpha}\xi^{A}D_{\beta}\xi_{A} 
-M_{A,B}\xi^{A}\xi^{B} \,],
\ee
where the {\it mass matrix} \cite{Callan:1989nz,Drukker:2000ep}
\be
M_{AB}=R_{AM,BN} t^{a M} t_{a}^{N}
\ee
is defined through two vectors tangent to the worldsheet
\be
t_{a}^{A}= E^{A}_{m} e_{a}^{\alpha}\partial_{\alpha} X^{m}\,,
\ee
and the covariant derivative now reads
\be
D_{\alpha}\xi^{A}=\partial_{\alpha} X^{A}+\Omega^{A}_{\ \  B\, m} \xi^{B}\partial_{\alpha} X^{m},
\ee
where the spin-connection  $\Omega^{A}_{\ \  B m}$ replaced the usual  Christoffel symbol.
To better understand the geometrical structure of the Lagrangian \eqref{actionDGT}, we 
introduce $(D-2)$
orthonormal vector fields $N_{i}^{A}$   orthogonal to the worldsheet, and 
decompose the field $\zeta^{A}$
in  directions tangent ($x^a$) and orthogonal ($y^i$)  to it
\be
\xi^{A}= x^{a} t_{a}^{A}+y^{i} N_{i}^{A}\ \ \ \  {a=1,2}\ \ \ \  {\rm and}\ \ \  {i,j,k,l=1,\dots,D-2}.
\ee
As is well-known from the general theory of submanifolds \cite{opac-b1121838},
this  decomposition  carries over to the covariant derivatives and one finds
\be
\label{extrin}
t_{A}^{a} D_{\beta} \xi^{A}=\mathcal{D}_{\beta}x^{a}-K^{a}_{A\beta} N^{A}_{i} 
y^{i}\,,\qquad\qquad
N^{i}_{ B} D_{\beta} \xi^{B}= \mathcal{D}_{\beta} y^{i}+x^{a} N^{i}_{B} K^{B}_{a\beta}\,.
\ee
Here $\mathcal{D}_\alpha$ is the covariant derivative on the worldsheet and it acts differently  on $x^a$ and  $y^i$
\be
\label{covder}
\mathcal{D}_\alpha x^a\equiv \partial_\alpha x^a +\omega^a{}_{b\,\alpha} x^b\ \ \ \ \mathrm{and}\ \ \ \ 
\mathcal{D}_\alpha y^i \equiv \partial_\alpha y^i -A^i{}_{j\alpha} y^j,
\ee
since $x^a$ lives in the tangent bundle  of the worldsheet, while $y^i$ is a section of the normal bundle.  The
connection $A^i{}_{j\alpha}$ on the normal bundle\footnote{The normal bundle is an $SO(D-2)$ bundle and  $A^i{}_{j\alpha}$ is a gauge connection induced on this bundle by the classical solution.} is given  by 
\be\label{normalconnection}
 A^{i}{}_{j\beta}\equiv N^{jB} D_\beta N^{i}_B= N^{B}_j(\partial_{\beta} N^{i}_{B}-N^{i}_{C}\,\Omega^{C}_{\ \ B\beta}).
\ee 
As usual the action of $\mathcal{D}_\alpha$ on tensors  with indices on both bundles is obtained combining  the two actions in  \eqref{covder}. The
tensor $K^a_{A\beta}=E_{Am} e^{a\alpha} K^m_{\alpha\beta}$ in  \eqref{extrin} is the extrinsic curvature  \eqref{extrinsic} of the embedding  expressed in a mixed basis.
In the following we will make use  of  the {\it Gauss-Codazzi} equation
\be
\label{Gauss1}
R_{ACBD} t^{A}_{\alpha} t^{C}_{\rho} t^{B}_{\beta} t^{D}_{\sigma}= ^{(2)}\!\!R
_{\alpha\rho\beta\sigma}+\eta_{AB}K^{A}_{\rho\beta} K^{B}_{\sigma\alpha}-\eta_{AB}K^{A}_{\rho\sigma} K^{B}_{\beta\alpha}\,,
\ee
an integrability condition relating the curvature $^{(2)}\!\!R_{\alpha\rho\beta\sigma}$ to the extrinsic and background geometry as characterized by the extrinsic curvature $K^{A}_{\alpha\beta} $ and the space-time Riemann tensor $R^A_{\phantom{A}CBD}$. Another useful constraint on the covariant derivative of the extrinsic curvature is provided by the {\it Codazzi-Mainardi} equation
\be\label{codazzi}
\mathcal{D}_{\alpha} K^{i}_{\beta\gamma}-\mathcal{D}_{\beta} K^{i}_{\alpha\gamma}=R_{MNRS} t^{M}_{\alpha}
t^{N}_{\beta}t^{S}_{\gamma} N^{R i}\, \  \  \ \ \ \  [K^i_{\alpha\beta}\equiv K^A_{\alpha\beta} N^{i}_{A}]~. 
\ee
Taking into account  \eqref{Gauss1}, \eqref{codazzi} and the  equation of motion \eqref{meanvanish} for the  background,
 the quadratic Lagrangian \eqref{actionDGT} finally appears to be
\be\label{lagrinterm}
\begin{split}
\mathcal{L}=\sqrt{h} &\bigl [(h^{\alpha\beta} \mathcal{D}_{\alpha} x^{a} \mathcal{D}_{\beta} x_{a}-^{(2)}\!\!R_{a b} x^{a} x^{b})+
h^{\alpha\beta} \mathcal{D}_{\alpha} y^{i} \mathcal{D}_{\beta} y_{i}-\\
&-2
h^{\alpha\beta}(\mathcal{D}_{\alpha}x^{a}K_{i, a\beta} y^{i}-\mathcal{D}_{\alpha}y^{i}x^{a}K_{i, a\beta} )- 2m_{a i} x^{a} y^{i}-m_{ij} 
y^{i}y^{j}\bigr]\,.
\end{split}
\ee
Above, 
the matrices appearing in the mass terms are defined as follows
\be
m_{ai}=
 -h^{\alpha\beta}\nabla_{\alpha} K_{i,a\beta}
 \ \ \ \ {\rm and}\ \ \ \  m_{ij}=R_{AM,BN} t^{c M} t_{c}^{N} N^{A}_{i}
 N^{B}_{j}-h^{\alpha\beta} h^{\rho\sigma}K_{i,\alpha\rho}  
K_{j,\beta\sigma}.
\ee
So far  we have treated the independent metric $h_{\mu\nu}$ as  a non-dynamical field. 
We should recall that in Polyakov's formulation also $h_{\mu\nu}$  fluctuates, 
\be
h_{\mu\nu}= \hat{h}_{\mu\nu}+ \text{\textfrc{h}}_{\mu\nu},
\ee
with respect to a classical background $\hat 
h_{\rho\sigma}=e^{\varphi}\gamma_{\rho\sigma}$
which solves the equations of motion. In particular this means that all the $h_{\mu\nu}$ appearing in the previous analysis must be
replaced with $\hat 
h_{\rho\sigma}$.
The quadratic part of the Polyakov action involving the fluctuations $\h_{\mu\nu}$ 
pedantically reads
\be
\begin{split}\label{Sdeltahp}
S_{\hh}=
&\int d^{2}\sigma e^{\varphi}\sqrt{\gamma}  \,\big[\,\textstyle{\frac{1}{4}}\,
( \gamma^{\sigma\mu} \gamma^{\rho\nu}+
 \gamma^{\sigma\nu} \gamma^{\rho\mu}-\gamma^{\mu\nu}\gamma^{\rho\sigma}\,)
 \h_{\mu\nu}  \h_{\rho\sigma}-\\
 &-(\,\mathcal{D}^{\mu}x^{\nu}+\mathcal{D}^{\nu} x^{\mu}-\gamma^{\mu\nu} \mathcal{D}_{\alpha}x^{\alpha}
 -2 y_{i} K^{i\mu\nu}\,)\h_{\mu\nu}\big],
 \end{split}
\ee
where $x^\alpha= e_a^\alpha x^a$. Of course \eqref{Sdeltahp} only depends on the traceless part $\bar \h_{\mu\nu}$ of $\h_{\mu\nu}$ as
required by Weyl invariance.
To deal with the quadratic fluctuation \eqref{lagrinterm} and \eqref{Sdeltahp}, we have different possibilities.
For instance, we can  choose the 
\emph{conformal gauge} for the metric fluctuations
\be\label{confgauge}
\h_{\mu\nu}=e^{\varphi}\gamma_{\mu\nu}\delta\varphi.
\ee
The action $S_{\hh}$ then vanishes identically, while 
the  ghost action associated to the choice \eqref{confgauge} is 
\be
\mathcal{L}_{\rm ghost}=\frac{1}{\sqrt{2} }b^{\mu\nu} \delta^{\rm diff.}_{c}\left (\h_{\mu\nu}-\frac{1}{2} \gamma_{\mu\nu}\gamma^{\alpha\beta}\h_{\alpha\beta}\right)=\frac{1}{\sqrt{2}}
b^{\mu\nu} (\mathcal{D}_{\mu} c_{\nu}+\mathcal{D}_{\nu} c_{\mu}-\gamma_{\mu\nu}\mathcal{D}_{\alpha} c^{\alpha}).
\ee
Here $\delta^{\rm diff.}_{c}$ is the BRST variation under diffeomorphism  of $h_{\mu\nu}$ with parameter $c$, and the ghost $b^{\mu\nu}$ is a symmetric traceless tensor. 
The full ghost contribution is therefore encoded into a functional determinant, obtained by integrating over $c$ and $b^{\mu\nu}$. It will correct the one-loop quantum result of the bosonic fluctuations, as we will see in the following. Concretely the computation of the ghost determinant means here to solve the following eigenvalue problem in the background geometry
\be
\left\{\begin{split}
&\frac{1}{\sqrt{2}}(\mathcal{D}_{\mu} c_{(n)\nu}+\mathcal{D}_{\nu} c_{(n)\mu}-\gamma_{\mu\nu}\mathcal{D}_{\alpha} c_{(n)}^{\alpha})=\lambda_{n} b_{(n)\mu\nu}\\
&-\sqrt{2}\mathcal{D}_{\mu} b^{\mu\nu}_{(n)}=\lambda_{n} c^{\nu}_{(n)}.
\end{split}\right.
\ee
From the first equation for $b_{\mu\nu}$  one gets
\be
-(\square c_{(n)}^{\nu}+R^{\nu}_{\ \mu} c_{(n)}^{\mu})=\lambda^{2}_{n} c^{\mu}_{(n)}
\ee
and therefore the ghost determinant is 
\be\label{detghost}
\triangle_{gh}=\prod_{n}\lambda_{n}=\left(\prod_{n}\lambda^{2}_{n}\right)^{1/2}=
\left[\det(-\square\delta^{\nu}_{\mu}-R^{\nu}_{\ \mu})\right]^{1/2}\,.
\ee
We can now decouple the longitudinal fluctuation $(x^a)$ from the transverse ones $(y^i)$. We start
from the action \eqref{lagrinterm},  
and  derive the equation of motion for the fluctuation parallel to the 
worldsheet: 
\be
\label{w1}
\square x_{\alpha} +R_{\alpha\beta} x^{\beta}=\mathcal{D}^{\beta} 
B_{\beta\alpha}\,,\qquad\qquad B_{\alpha\beta}=2 y_{i} K^{i}_{\alpha\beta}
\ee
where we introduced the traceless tensor  $B_{\alpha\beta}$. 
 This equation can be equivalently written  as follows
\be\label{eqforT}
P_1(x)_{\alpha\beta}=\mathcal{D}_{\beta} x_{\alpha}+\mathcal{D}_{\alpha} x_{\beta}-\gamma_{\alpha\beta}
\mathcal{D}_{\rho} x^{\rho} =B^{\parallel}_{\alpha\beta}
\ee
where we have introduced a projector $P_1$, acting on the space of vectors and producing symmetric traceless tensors. We can conveniently decompose the traceless symmetric tensor $B_{\alpha\beta}$ into $B^{\parallel}_{\alpha\beta}+B^{\perp}_{\alpha\beta}$,
with $B^{\parallel}_{\alpha\beta}\in \mathrm{range}(P_1)$ and $B^{\perp}_{\alpha\beta}\in\mathrm{range}(P_1)^\perp=\mathrm{Ker }(P_1^\dagger)$.  We remark that we will be only interested in worldsheet with the topology of the sphere or the disc, where $B^{\perp}_{\alpha\beta}=0$ and thus $B_{\alpha\beta}=B^{\parallel}_{\alpha\beta}$\footnote{This fact corresponds to the the well-known property of the absence of non-trivial Beltrami differentials at genus 0.}.
Considering now  a solution $\bar x$ of  \eqref{eqforT} and performing the  shift 
$
x^{a}\mapsto \bar x^{a}+ x^{a}~
$
in the  path-integral, all  mixed terms $ x y$ in \eqref{lagrinterm} disappear and an additional contribution shows up in the quadratic action
\be
\bar x^{\alpha} \mathcal{D}^{\beta} B_{\alpha\beta}=\mathcal{D}^{\beta}(\bar x^{\alpha} B_{\alpha\beta})-
\frac{1}{2}B^{\alpha\beta}B_{\alpha\beta}=
-2 y_{i} y_{j} K^{i}_{\alpha\beta} K^{j\alpha\beta}\,.
\ee 
We are left therefore  with the quadratic Lagrangian
\be
\mathcal{L}\equiv \mathcal{L}_{\rm long}+ \mathcal{L}_{\rm transv}
\ee
where
\be\label{Llong}
\mathcal{L}_{\rm long}=\sqrt{\gamma}  (\gamma^{\alpha\beta} \mathcal{D}_{\alpha} x^{a} \mathcal{D}_{\beta} x_{a}-^{(2)}\!\!R_{a b} x^{a} x^{b})
\ee
and
\be
\begin{split}
\label{L1}
\mathcal{L}_{\rm transv} &=  \sqrt{\gamma}(\gamma^{\alpha\beta} \mathcal{D}_{\alpha} y^{i} \mathcal{D}_{\beta} y_{i}-\mathcal{M}_{ij} y^{i} y^{j})\,, \\
\mathcal{M}_{ij} &= R_{AM,BN} t^{c M} t_{c}^{N} N^{A}_{i}
 N^{B}_{j}+K_{i,\alpha\beta}  
K_{j}^{\alpha\beta}.
\end{split}
\ee
%
After the above redefinition,
the operator controlling the fluctuations $x^a$ parallel to the worldsheet  in \eqref{Llong} coincides with the one appearing in the ghost determinant 
\eqref{detghost}, but we remark that  this does \emph{not} mean in general that the 
corresponding determinant will be simply cancelled by the ghost contribution. For instance in the case of open strings   
different boundary conditions should be imposed for the two determinants. Moreover  the treatment of the ghost operator requires 
additional care since it might contain zero modes associated to the Killing vectors of the worldsheet metric $\gamma_{\alpha\beta}$.

\noindent
We can reach the same  expression for $\mathcal{L}_{\rm transv}$ following an alternative way. Instead of setting to zero the  traceless  part $ \bar \h_{\mu\nu}$ of $\h_{\mu\nu}$
by choosing  the conformal gauge fixing \eqref{confgauge}, we  
can decouple it   from the fluctuations of the embedding coordinates through 
the following shift
\be
 \bar \h_{\mu\nu}\mapsto \bar \h_{\mu\nu}+\mathcal{D}_{\mu}x_{\nu}+\mathcal{D}_{\nu} x_{\mu}-\gamma_{\mu\nu} \mathcal{D}_{\beta}x^{\beta}-2 y_{i} K^{i}_{\mu\nu}.
\ee
We find a quadratic algebraic  action for $ \bar \h_{\mu\nu}$ and an additional term
leading to the bosonic Lagrangian 
\be
\label{cippa}
\begin{split}
\mathcal{L}
=\frac{1}{2}\sqrt{\gamma}\gamma^{\alpha\beta}\gamma^{\rho\sigma}\bar \h_{\alpha\rho} \bar \h_{\beta\sigma}&+
\sqrt{\gamma} \bigl [
\gamma^{\alpha\beta} \mathcal{D}_{\alpha} y^{i} \mathcal{D}_{\beta} y_{i}-(m_{ij}+2K_{i\mu\nu} K^{\mu\nu}_j)y^{i} y^{j}\bigr]+\\
&
+2\sqrt{\gamma}
\gamma^{\alpha\beta}(\mathcal{D}_{\alpha}x^{a}K_{i, a\beta} y^{i}+\mathcal{D}_{\alpha}y^{i}x^{a}K_{i, a\beta} +\gamma^{\alpha\beta}\mathcal{D}_{\alpha} K_{i,a\beta} x^{a} y^{i})\,,
\end{split}
\ee
where  the last line is a total divergence and can be neglected. The Lagrangian for the transverse fluctuations is again given by \eqref{L1}.
We notice that in eq. \eqref{cippa} the longitudinal fluctuations have completely disappeared from the Lagrangian. This second way to proceed is similar to  a sort of  {\it static 
gauge fixing} (where longitudinal fluctuations are taken to vanish).  More precisely, in the present approach the longitudinal degrees of freedom of the metric decouple,  and their role in the Lagrangian  is taken by the traceless part of the metric fluctuation, which however possesses an algebraic gaussian action. 
If we integrate it out, we find an ultra-local functional determinant, whose  careful regularisation and evaluation   hides some 
of  the subtleties  discussed  below eq. \eqref{detghost}.

\noindent
Another option to get rid of the longitudinal fluctuation is to  choose the static gauge fixing $x^a=0$.  Then   \eqref{lagrinterm}
reduces to a Lagrangian  for the transverse fluctuation only, with a mass matrix given not by $\mathcal{M}_{ij}$ but by $m_{ij}$.  However
we  must recall that the  transverse fluctuation are still coupled to the  metric fluctuations (see eq. \eqref{Sdeltahp}):
\be
\begin{split}\label{Sdeltah}
S_{\hh}=
&\int d^{2}\sigma e^{\varphi}\sqrt{\gamma}  \,\big[\,\textstyle{\frac{1}{2}}\,
\gamma^{\sigma\mu} \gamma^{\rho\nu}
\bar \h_{\mu\nu}  \bar \h_{\rho\sigma}+
 2 y_{i} K^{i\mu\nu} \bar \h_{\mu\nu}\big]~.
 \end{split}
\ee
If we again  eliminate the metric fluctuation $\bar \h_{\mu\nu}$ through its equation of motion, we get back to the mass matrix $\mathcal{M}_{ij}$
and to $\mathcal{L}_{\rm transv}$  in \eqref{L1}. 

\noindent
We remark finally that starting from the Nambu-Goto action, where no dynamical worldsheet metric is present, in the static gauge $x^a=0$ we would directly obtain $\mathcal{L}_{\rm transv}$.

\bigskip

\noindent
In view of the analysis above, we will focus our attention on transverse fluctuations and will examine closely  the structure of the Lagrangian  in \eqref{L1}
which governs their dynamics. These modes are in general coupled between themselves and  we can distinguish  two different sources of coupling: the  
$\grSO(D-2)$ gauge connection  $A^{i}{}_{j\alpha}$ induced on the normal bundle and the mass matrix   $\mathcal{M}_{ij}$. In the following we shall
discuss some general properties of these two objects.


 \subsection{The normal bundle}
 \label{normalbundle}
Let us begin to discuss the geometric structure of the normal bundle. The curvature $F^{i}{}_{j \alpha\beta}\equiv\partial_\alpha A^{i}{}_{j\beta}-\partial_\beta A^{i}{}_{j\alpha}+A^{i}{}_{\ell\alpha} A^{\ell}{}_{j\beta}-A^{i}{}_{\ell\beta} A^{\ell}{}_{j\alpha}$ can be easily evaluated in terms of the Riemann curvature of the target space 
 and of the extrinsic curvature  through the  {\it Ricci equation:}
 \be
 \label{NormalCurvature}
 F^{i}{}_{j\alpha\beta}=-R_{ABMN} t^{A}_{\alpha} t^{B}_{\beta} N^{iM} N^{N}_j-h^{\rho\sigma} (K^{i}_{\rho\alpha}K_{j\sigma\beta}-K^{i}_{\rho\beta}K_{j\sigma\alpha}).
 \ee
 Focussing on the case  of $\ads^5\times \sphere^5$, we see that the contribution of the Riemann tensor to the curvature of the normal bundle vanishes identically as in flat space. 
 In fact the Riemann tensor for this background is  given by
\be\label{riemannmaxsusy}
R_{AB,CD}=-(\hat P_{AC} \hat P_{BD}-\hat P_{AD} \hat P_{BC})+(\bar P_{AC} \bar P_{BD}-\bar P_{AD} \bar P_{BC}),
\ee
where $\hat P_{AB}$ is the projector on $\ads_{5}$ and $\bar P_{AB}$ is the on $\sphere^{5}$. Obviously the sum of  the two projectors gives the identity:
\be
\hat P_{AB}+\bar P_{AB}= \eta_{AB}.
\ee
The  contractions $R_{ABMN} t^{A}_{\alpha} t^{B}_{\beta} N^{iM} N^{N}_j$ can be written as
\be
\label{contraction1}
R_{ABMN} t^{A}_{\alpha} t^{B}_{\beta} N^{iM} N^{N}_j\!\!=\!\![(\bar{t}_\alpha\cdot \bar{N}^i)(\bar{t}_\beta\cdot \bar{N}_j)-(\bar{t}_\beta\cdot \bar{N}^i)(\bar{t}_\beta\cdot
\bar{N}_j)]\!
-\![(\hat{t}_\alpha\cdot \hat{N}^i)(\hat{t}_\beta\cdot \hat{N}_j)-(\hat{t}_\beta\cdot \hat{N}^i)(\hat{t}_\beta\cdot
\hat{N}_j)],
\ee
where we introduced the inner product $(a\cdot b)\equiv \eta_{CD} \,a^C \,b^D$ as a product over flat spacetime indices.
Above, the  hats and the bars over the vectors denote the projection of the vectors on $\ads_5$ and $\sphere^5$ respectively,
namely
\be
\label{hatbar}
\hat V^A= \hat P^A{}_B V^B\ \ \ \ \ \ \mathrm{and}\ \ \ \ \ \  \bar V^A= \bar P^A{}_B V^B.
\ee
We can now use that 
\be
\label{ortotN}
(\bar{t}_\alpha\cdot \bar{N}^i)=(\bar{t}_\alpha\cdot {N}^i)=((t_\alpha-\hat t_\alpha)\cdot {N}^i)=-(\hat t_\alpha \cdot {N}^i)=-(\hat t_\alpha \cdot \hat{N}^i),
\ee
and we remain with the following expression for the normal curvature
\be
F^{i}{}_{j\alpha\beta}=-\gamma^{\rho\sigma} (K^{i}_{\rho\alpha}K_{j\sigma\beta}-K^{i}_{\rho\beta}K_{j\sigma\alpha})=(\gamma^{\rho\sigma} \epsilon^{\mu\nu}K^{i}_{\rho\mu}K_{j\sigma\nu})\epsilon_{\alpha\beta}\equiv \sqrt{\gamma} \epsilon_{\alpha\beta}
\mathcal{F}^{i}{}_{j}
\ee 
which also holds in flat space.  The normal bundle is  {\it   flat}  when $F^{i}{}_{j,\alpha\beta}=0$. 
In that case, we can always choose the normal vectors $N^i$ so that $A^i{}_{j\alpha}=0$  and the covariant derivative acting on the transverse fluctuations reduces to the usual one.
This occurs, for instance,  when the minimal surface is confined in a three-dimensional subspace of our target space: the extrinsic curvature is in fact not vanishing just in one normal direction~\cite{Kavalov:1986nx,Sedrakian:1986np,Langouche:1987mw,Langouche:1987my,Langouche:1987mx,Karakhanian:1990nd, Drukker:2000ep}.

\noindent
For a generic worldsheet, which solves the equation of motion in the $\ads_5\times \sphere^5$ background,  the extrinsic curvarture  defines, at most,  two independent vector fields normal to the worldsheet\footnote{This follows from the fact that  we have just two independent components of the extrinsic curvature ({\it e.g.} $K^\ell_{11}$ and  $K^\ell_{12}$). Alternatively, we can argue this result, in a covariant way, from the following matrix relation satisfied by $\cal F$:
$$
{\cal F}^4=\frac{1}{2}\mathrm{Tr}(\mathcal{F}^2) {\cal F}^2.
$$ }.
 In fact the matrix $\mathcal{F}^i{}_{j}$ can be always put in the
form 
\be\label{Fij}
\mathcal{F}^i{}_{j}=(k^i  h_j- k_j h^i),
\ee
where $(h^i\cdot t_\alpha)= (k^i\cdot t_\alpha)=0$ and we can choose $(k\cdot h)=0$ without loss of generality.

 \subsection{Mass matrix and sum rules}
 \label{sec:mass_matrix_sum_rules}
The next step is to examine more closely 
the mass matrix \eqref{L1} for  the transverse bosonic  degrees of freedom. For a   generic classical  background 
it has the form
\be
\label{mij2}
\mathcal{M}_{ij}=R
_{AM,BN} t^{c M} t_{c}^{N} N^{A}_{i}
 N^{B}_{j}+K_{i,\alpha\beta}  
K_{j}^{\alpha\beta}.
\ee
There are few  general properties of $\mathcal{M}_{ij}$  that can be easily read  from \eqref{mij2}, since the
 embedding equations for a sub-manifold do not provide a direct constraint on the contraction $R
_{AM,BN} t^{c M} t_{c}^{N} N^{A}_{i} N^{B}_{j}$.
However, its  trace $\mathrm{Tr}(\mathcal{M})$ admits a quite simple and compact expression in  terms 
of geometric  quantities. If we use the  completeness relation
\be
N_i^A N^{iB}= \eta^{AB}-t_c^A t^{cB},
\ee
we can rewrite the trace of \eqref{mij2} as follows
\begin{equation}\label{rop1}
\mathrm{Tr}({\cal M})
=R_{MN} t^{c M} t_{c}^{N}-
R_{AM,BN} t^{c M} t_{c}^{N}t_d^A t^{dB}+\mathrm{Tr}(K^2)\,,
\end{equation}
where
\be
\mathrm{Tr}(K^2)=\gamma^{\alpha\beta}\gamma^{\rho\sigma}\eta_{AB}K^{A}_{\rho\beta} 
K^{B}_{\sigma\alpha}~.
\ee
By mean of the  Gauss equation \eqref{Gauss1} we can reduce the second term in \eqref{rop1} to
\begin{equation}\label{riemannsummed}
R_{AM,BN} t^{c M} t_{c}^{N}t_d^A t^{dB}= ^{(2)}\!\!R+\mathrm{Tr}(K^2)\,,
  \end{equation} 
 where $^{(2)}\!\!\,R$ is the two-dimensional (intrinsic) scalar curvature.
This  leads to the sum rule
\begin{equation}\label{trace}
 \mathrm{Tr}({\cal M})=R_{MN}\,t^{cM}t^N_c -  ^{(2)}\!\!R
\end{equation}
in terms of the Ricci tensor $R_{MN}$ and of the two-dimensional  curvature. In the case of Einstein spaces 
(where $R_{MN}= n \,G_{MN}$) 
this gives  $\mathrm{Tr}({\cal M})=2 n-^{(2)}\!\!R$. 

\bigskip
\noindent
We now particularize the analysis of the 
mass matrix  \eqref{mij2}  for a string moving in 
 $\ads_{5}\times \sphere^{5}$ background. 
 The Riemann tensor for the background $\ads_5\times \sphere^5$ is  given in term of projectors by \eqref{riemannmaxsusy}.
The first contribution to the mass matrix takes the form
\be
\label{M1}
\begin{split}
\gamma^{\alpha\beta}R_{ARBS} t^{A}_{\alpha} t^{B}_{\beta} N_i^R N_j^S  =&  
-\gamma^{\rho\sigma} ( \hat t_{\rho}\cdot\hat t_{\sigma} )  (\hat N_i\cdot \hat N_j)+ \gamma^{\alpha\beta} 
(\hat N_i\cdot \hat t_{\alpha} )(\hat N_j \cdot\hat t_{\beta})
+\\
&+\gamma^{\rho\sigma} (\bar t_{\rho} \cdot \bar t_{\sigma} )\,(\bar N_i \cdot\bar N_j)
- \gamma^{\alpha\beta} (\bar N_i\cdot \bar t_{\alpha})(\bar N_j \cdot\bar t_{\beta}),
\end{split}
\ee
where the  hats and the bars over the vectors again  denote the projection of the vectors on 
$\ads_5$ and $\sphere^5$ respectively and they are defined in \eqref{hatbar}. The relation 
\eqref{ortotN} easily shows that the mixed terms in  \eqref{M1} cancel and we obtain
\be
\label{M1}
\begin{split}
\gamma^{\alpha\beta}R_{ARBS} t^{A}_{\alpha} t^{B}_{\beta} N_i^R N_j^S  =&  
-\gamma^{\rho\sigma} ( \hat t_{\rho}\cdot\hat t_{\sigma} )  (\hat N_i\cdot \hat N_j)+
\gamma^{\rho\sigma} (\bar t_{\rho} \cdot \bar t_{\sigma} )\,(\bar N_i \cdot\bar N_j).
\end{split}
\ee
With a small  abuse of language we shall introduce the following notations
\be\label{massesads5s5}
m^2_{AdS_5}\equiv  \gamma^{\rho\sigma} ( \hat t_{\rho}\cdot\hat t_{\sigma} )\  \ \ \ \ \  \mathrm{and}\ \ \ \ \ \ 
m^2_{S_5}\equiv -\gamma^{\rho\sigma} (\bar t_{\rho} \cdot \bar t_{\sigma} ).
\ee
Then the complete mass matrix is  given by
\be
\label{Mtot}
\begin{split}
\mathcal{M}_{ij} =&  -m^2_{AdS_5}  (\hat N_i\cdot \hat N_j)-m^2_{S^5}\,(\bar N_i \cdot\bar N_j)+K_{i\alpha\beta} K_{j}^{\alpha\beta}.
\end{split}
\ee
It is worth noticing that the two scalar quantities   $m^2_{AdS_5}  $ and $m^2_{S^5}$ can be computed  in terms of the
two-dimensional scalar curvature $^{(2)}\!\!\,R$ and $\mathrm{Tr}(K^2)$. We first observe that
\be
\label{S5-V-AdS5}
m^2_{AdS_5}-
m^2_{S^5}=\gamma^{\rho\sigma} (\bar t_{\rho}\cdot\bar t_{\sigma} )+\gamma^{\rho\sigma} ( \hat t_{\rho}\cdot\hat t_{\sigma} )=\gamma^{\rho\sigma} (t_\rho\cdot t_\sigma)=2,
\ee
while another relation between these two quantities can be obtained from  Gauss equation  \eqref{Gauss1} in its contracted form. After some simple manipulations we end up with
\begin{align}
\label{Gauss}
 ^{(2)}\!\!R+\mathrm{Tr}(K^2)=&\gamma^{\alpha\beta}\gamma^{\rho\sigma}
R_{ACBD} t^{A}_{\alpha} t^{C}_{\rho} t^{B}_{\beta} t^{D}_{\sigma}=
m_{S_5}^4
-m_{AdS5}^4+(\hat t_{c}  \cdot \hat t^{d})   (\hat t_d \cdot \hat t^{c})-(\bar t_{c}  \cdot \bar t^{d})   (\bar t_d \cdot \bar t^{c})=\nn\\
=&-2 (m^2_{S_5}+1)~.
\end{align}
We finally obtain:
\be\label{ms5mads5}
m^2_{AdS_5}=-\frac{1}{2}(^{(2)}\! R+\mathrm{Tr}(K^2))+1\,\qquad\mathrm{and}\qquad m^2_{S_5}
=-\textstyle{\frac{1}{2}}\,\big(^{(2)}\! R+\mathrm{Tr}(K^2)\,\big)-1.
\ee
The results \eqref{ms5mads5} simplify the explicit evaluation of the trace of the mass matrix $\mathrm{Tr}({\cal M})$ for the 
case of $\ads_5\times \sphere^5$. Since 
$
R_{MN}=-4\,\hat P_{MN}+4\bar P_{MN}$,
we find 
\be
\label{sumrulebos}
\begin{split}
\mathrm{Tr}({\cal M})=& R_{MN} t^{c M} t_{c}^{N}-^{(2)}\!\!R=-4(m^2_{S^5}+m^2_{AdS_5})-  ^{(2)}\!\!R=4\,\big(^{(2)}\!\!R+\mathrm{Tr}(K^2)\,\big)-  ^{(2)}\!\!R=\\
=& 3\,^{(2)}\!R+4\,\mathrm{Tr}(K^2)~.
\end{split}
\ee

\noindent
The structure of the mass matrix  \eqref{Mtot} can be further constrained assuming particular properties for the classical background.
The simplest geometrical configuration  is when in 
\eqref{Mtot} either $\hat t_c=0$ or $\bar t_c=0$ ($c=1,2$).  In that case 
the minimal surface $\Sigma$ is confined in one of the two spaces:  $\ads_5$ if  $\bar t_c=0$ or $S^5$  if $\hat t_c=0$. 
Let us focus on the first possibility; the second one can be discussed in complete analogy.  The mass matrix then reduces to
\be
\begin{split}
\mathcal{M}_{ij} =&  
-2 (\hat N_i\cdot \hat N_j)+K_{i\alpha\beta} K_{j}^{\alpha\beta},
\end{split}
\ee
where we have used that  $m^2_{S^5}=(\bar t_c \cdot \bar t^c)=0$ and $m^2_{AdS_5}=m^2_{AdS_5}-
m^2_{S^5}=2$. The extrinsic curvature $K_{i\alpha\beta}$ is different from zero only for orthogonal directions lying in $\ads_5$.   Therefore we have $5$ massless scalar (${\frak m}_i=m_{S^5}=0\  \ \ [i=1,\dots,5]$),  one for each direction of $\sphere^5$. We 
can always choose a sixth direction (lying in $\ads_5$) orthogonal  to $\Sigma$ and to the  two  normal directions defined by $K_{i\alpha\beta}$. The mass  ${\frak m}_6$ of this sixth   scalar is  
\be
{\frak m}_6^2=-2.
\ee
Finally, we have to select the last two orthogonal  directions (i=7,8)  and  we choose
the only two orthonormal eigenvectors  of $K_{i\alpha\beta} K_{j}^{\alpha\beta}$ with non vanishing eigenvalues. They always exist if the normal bundle is not flat. Then the two masses are given by
\be
{\frak m}^2_7=\lambda_1-2 \ \ \mathrm{and} \ \   {\frak m}^2_8=\lambda_2-2.
\ee
Here $\lambda_1$ and $\lambda_2$ are the two non-vanishing eigenvales of $K_{i\alpha\beta} K_{j}^{\alpha\beta}$ and they are determined in terms of the geometric quantity of the surface through the relations:
\be
\label{eigen}
\lambda_1+\lambda_2=\mathrm{Tr}(K^2)=-{}^{(2)}R-2\ \ \  \ \mathrm{and}\ \ \   \lambda_1\lambda_2=
\frac{1}{2}[ (\mathrm{Tr}(K^2))^2 -\mathrm{Tr}(K^4)]=\frac{1}{2} \mathrm{Tr}(\mathcal{F}^2),
\ee
where $\mathrm{Tr}(K^4)=K^i_{\alpha\beta} K_{j}^{\alpha\beta} K^{j}_{\rho\sigma}K_{i}^{\rho\sigma}$.
If the normal bundle is flat, $\mathcal{F}=0$ and one of the two eigenvalues vanishes, {\it e.g.}  $\lambda_1=0$. Then the two masses collapse to the known result \cite{Drukker:2000ep,Drukker:2011za} 
\be\label{flatbundlemasses}
{\frak m}^2_7=-2 \ \ \mathrm{and} \ \   {\frak m}^2_8=-{}^{(2)}\!R-4.
\ee
Let us turn our attention  to the general case where the worldsheet extends {both} in  $\ads_5$ and $\sphere^5$ spaces
and the mass matrix has the general form  \eqref{Mtot}.  The first step is to  choose  two of the fluctuations ($i=7,8$) along 
the two orthogonal directions ($h$ and $k$) with non-vanishing extrinsic curvature. These two directions  are defined up to
a rotation in  $(h,k)-$plane. We fix this freedom by  choosing   $h$ and $k$ to be 
the only two orthonormal eigenvectors  of $K_{i\alpha\beta} K_{j}^{\alpha\beta}$ with non vanishing eigenvalues.
 Then the only non vanishing component of the field strength in the normal bundle is $\mathcal{F}^7{}_8$ as discussed
 in subsec. \ref{normalbundle}.
 
\noindent
The bosonic masses  can be  analysed in details if  the field strength $\mathcal{F}^7{}_8$ is essentially {\it abelian}, namely if the only component of the  connection different from zero is given by $A^7{}_8$. In this case the Codazzi-Mainardi equation \eqref{codazzi} implies for the normal directions  $  i=1,\dots, 6$
\be\label{codazzinormal}
\gamma_{\beta\gamma} (\bar t_{\alpha}\cdot N_{i}) -\gamma_{\alpha\gamma} (\bar t_{\beta}\cdot N_{i})
=\mathcal{D}_{\alpha} K^{i}_{\beta\gamma}-\mathcal{D}_{\beta} K^{i}_{\alpha\gamma}=0,
\ee
which immediately translates in 
\be
\label{orthogonality}
(\bar t_{\beta}\cdot N_{i})= (\hat t_{\beta}\cdot N_{i})=0\ \ \ \ \ \mathrm{for} \ \ i\ne 7,8.
\ee
We find that the remaining six  normal directions are orthogonal  both to  $\bar t_\alpha$ and  to $\hat t_\alpha$, implying 
that  some  of  these vectors completely lie in $\ads_5$, while the others in $\sphere^5$. Generically we expect to  find three of them in $AdS_5$ and three in $S^5$ (a different partition of the six vectors between the two subspaces may occur when some of $\hat t_\alpha$ or of $\bar t_\alpha$ vanishes).
 Because of the orthogonality relations \eqref{orthogonality},  the $8\times 8$  mass matrix $\mathcal{M}$
takes the form 
\be
\label{wmax}
\mathcal{M}=\begin{pmatrix} 
-m^{2}_{AdS_{5}} & 0 &0 &0&0&0&0&0\\
0 &-m^{2}_{AdS_{5}} &0 &0&0&0&0&0\\
0 &0&-m^{2}_{AdS_{5}}&0&0&0&0&0\\
0 &0&0&-m^{2}_{S^{5}}&0&0&0&0\\
0 &0&0&0&-m^{2}_{S^{5}}  &0&0&0\\
0 &0&0&0&0&-m^{2}_{S^{5}} &0&0\\
0 &0&0&0&0&0&m_{77}&m_{78}\\
0 &0&0&0&0&0&m_{87}&m_{88},
\end{pmatrix}
\ee
where  $m_{AdS_5}^2$ and $m_{S^5}^2$ are given in \eqref{ms5mads5}.  The  trace condition \eqref{trace} still constrains $m_{77}$ and $m_{88}$ and it yields that
\be
\label{fgh}
m_{77}+m_{88}= \mathrm{Tr}(K^2). 
\ee
With the help of  \eqref{fgh}, the reduced mass matrix in  the directions $7$ and $8$ can be cast into to the form
\be\label{reduced}
{\cal M}_{\rm red.}
\equiv \begin{pmatrix}
m_{77}&m_{78}\\
m_{87}&m_{88}
\end{pmatrix}=
\begin{pmatrix}
-m^2_{AdS_5} +2 (\bar N_7 \cdot\bar N_7)+\lambda_1 & 2  (\bar N_7 \cdot\bar N_8)\\
 2  (\bar N_7 \cdot\bar N_8) & m^2_{AdS_5} -2 (\bar N_7 \cdot\bar N_7)+\lambda_2
\end{pmatrix}
\ee
where $\lambda_1$ and $\lambda_2$ again obey \eqref{eigen}.

\noindent
In the general case, when $\mathcal{F}^7{}_8$ is not generated by only taking $A^7{}_8$ different from zero, the structure
of the mass matrix may become more  intricate. However we can always choose, at least,  two orthogonal  directions, one in $\sphere^5$ and one in $\ads_5$, which are orthogonal to the minimal surface $\Sigma$ and to the extrinsic curvature. The masses of these two fluctuations are then given by \eqref{ms5mads5}.

\Section{Fermionic fluctuations}
\label{sec:Fermionic fluctuations}

The full covariant GS string action in $\ads_5\times \sphere^5$ has a  complicated non-linear structure~\cite{Metsaev:1998it,Kallosh:1998zx}, but
to analyze the relevant fermionic contributions it is sufficient to consider only its quadratic part  here 
\begin{equation}\label{fermionGS}
L_{\rm 2F}=i\,\Big(\sqrt{\g}\g^{\alpha\beta}\,\delta^{IJ}-\ve^{\alpha\beta} s^{IJ}\big)\,\bar\theta^I\,\rho_\alpha\,D_\beta^{JK}\,\theta^K\,.
\end{equation}
Above, $\theta^I$ ($I=1,2$) are two ten-dimensional Majorana-Weyl spinors with the same chirality, $s^{IJ}=\textrm{diag}(1,-1)$, $\rho_{\alpha}$  are the worldsheet projections of the ten-dimensional Dirac matrices
 \be
 \rho_{\alpha}=E_{A m}\,\partial_{\alpha} \,X^{m}\,\Gamma^{A}\,,
 \ee 
 and $D_\alpha^{JK}$ is the two-dimensional pullback  $\partial_\alpha X^m\,D_m^{JK}$ of  the ten-dimensional  covariant derivative (here, flat and curved indices span the range from 1 to $D=10$),  
 sum of an ordinary spinor covariant derivative  and an additional ``Pauli-like'' coupling   to the RR flux background, $D_m^{JK}=\mathfrak{D}_m\,\delta^{JK}-\frac{1}{8\cdot 5!}F_{m_1...m_5}\Gamma^{m_1...m_5}\Gamma_m\,\epsilon^{JK}$. 

In the  $AdS_5\times S^5$ case,  it can be written as follows 
\begin{eqnarray}
\label{def_covariant_der}
&& D_\beta^{JK} \theta^K =
\mathfrak D_\beta^{JK}\theta^K+\mathcal F_{\beta}^{JK}\theta^K \,,\qquad\qquad 
\mathfrak D_\beta^{JK}= \delta^{JK} \left ( \p_\beta+ \quarter \p_\beta X^m \Omega_m^{AB} \Gamma_{AB} \right) \theta^K\,,
\ea
where  the flux term $\mathcal  F_{\beta}^{JK}$, responsible for the fermionic ``masses'', is~\footnote{
An alternative form  for the flux~\cite{Metsaev:1998it} is
\be\label{flux5p5}
\mathcal F_{\beta}^{JK} = -{i\over 2}\epsilon^{JK}\,\tilde\rho_\beta\,,\qquad\qquad \tilde\rho_\beta =  \Gamma_{A} \hat t^{A}_{\beta}+i\Gamma_{A} \bar t^{A}_{\beta} 
\ee 
with which the corresponding part of the gauge-fixed Lagrangian reads
\be\label{lagrfermflux}
L_{F}^{flux}=-\epsilon^{\alpha\beta}\,\bar\theta \,\rho_\alpha\,\tilde\rho_\beta\,\theta\,.
\ee 
Its equivalence with \eqref{fluxstar} and \eqref{lagrfermflux2} below is manifest in the $5+5$ basis of \cite{Metsaev:1998it}, see also discussion in~\cite{Drukker:2000ep}.
} 
\be
\label{fluxstar}
{{\mathcal F}}_{\beta}^{JK} =-{i\over 2}\epsilon^{JK} \Gamma_{\star} \rho_\beta\,, \qquad~~~ \Gamma_{\star} =i \Gamma_{01234}\,.
\ee
Looking for a general  formalism for fluctuations, there is a natural choice for the  $\kappa$-symmetry gauge-fixing that is
 viable in type IIB string action, where both Majorana-Weyl fermions in the GS action have the same chirality, namely~\footnote{A widely used alternative to \eqref{ksym} -- especially in the context of AdS/CFT -- is the light-cone gauge-fixing $\Gamma^+\theta^I=0$, where the light-cone might lie entirely in $\sphere^5$~\cite{Metsaev:2000yf,Metsaev:2000yu} or being shared between $\ads_5$ and $\sphere^5$~\cite{Callan:2003xr} (see~\cite{Arutyunov:2009ga} and further references therein). 
 One of the obvious advantages of the ``covariant'' gauge-fixing \eqref{ksym} is preservation of global bosonic symmetries of the action. A more general choice is $\theta_1=k\,\theta_2\,,\,\,\theta_2\equiv\theta$ where $k$ 
 is a real parameter whose dependence is expected to cancel in the effective action, see discussion in~\cite{Roiban:2007jf}. Yet another $\kappa$-symmetry gauge-fixing, albeit equivalent to \eqref{ksym}~\cite{Drukker:2000ep}, has been used for studying stringy fluctuations in $\ads_5\times \sphere^5$ in \cite{Forste:1999qn}.}
 \be\label{ksym}
 \theta^1= \theta^2\equiv \theta~.
\ee 
Since  $s^{11}=- s^{22}=1$, in the kinetic part of the gauge-fixed fermionic action only the term proportional to  $\g^{\alpha \beta}$ will survive after the $\kappa$-symmetry gauge-fixing, while in the flux part only the term proportional to $\ve^{\alpha\beta}$.

 \subsection{The fermionic kinetic term}
 \label{sec:fermions_kin}

 Let us first  focus on the  reduction of the ordinary spinor covariant derivative  $\mathfrak{D}_\beta$ in \eqref{def_covariant_der} and write  the relevant part of the $\kappa$-symmetry gauge-fixed Lagrangian as
\be\label{lagrfermkinRT}
L_{F}^{\rm kin}=2i\,\sqrt{\g}\g^{\alpha\beta}\,\bar\theta \,\rho_\alpha\,(\partial_\beta+\Omega_\beta)\,\theta\,,\qquad\qquad \Omega_\beta=\frac{1}{4}\,\Omega_\beta^{AB}\Gamma_{AB}\,.
\ee
The geometrical interpretation of the reduction procedure, already discussed  for the bosonic fluctuations, allows us to guess the final result
of this section. In fact  a straightforward, but tedious computation must yield, at the end,  the following form for the kinetic term:
\be\label{finalkinferm}
\begin{split}
\tilde{L}_F^{\rm kin}= 2i\,\sqrt{\g}\g^{\alpha\beta}\,\bar\Theta \, \Gamma^{a} \, e_{a\alpha}\, \mathcal{D}_\beta\,\Theta
\equiv2i\,\sqrt{\g}\g^{\alpha\beta}\,
\bar\Theta \, \Gamma^{a} \, e_{a\alpha}\, \big(\,\partial_{\beta}+\textstyle{\frac{1}{4}}\Gamma_{bc}\omega^{bc}_{\beta}-\textstyle{\frac{1}{4}} A^{\underline{i}\underline{j}}_{\beta} \Gamma_{\underline{i}\underline{j}}\,\big)\,\Theta \,.
\end{split}
\ee
Above, $\mathcal{D}_\beta$ is the two-dimensional covariant derivative acting on spinors, and it takes into account  that  $\Theta$ is now a two-dimensional spinor, which also transforms in the  spinor representation of the $\grSO(8)$ normal bundle. We remind that $e^{a}_{\alpha}$ denotes the worldsheet zweibein. The connection $A^{ij}_\beta$ is defined in \eqref{normalconnection}.  This pattern is  completely analogous to the one obeyed by the normal  bosonic fluctuations, which are 
scalars for the worldsheet, but vectors for the  normal bundle.

\noindent
We shall now see how the result  \eqref{lagrfermkinRT} emerges from the explicit analysis. Although we will work explicitly in $D=10$, we remark that the reduction of the canonical covariant derivative (passing from \eqref{lagrfermkinRT} to \eqref{finalkinferm}) is independent on the dimensionality of the spacetime, exactly as in the bosonic case~\footnote{The dimensionality of space-time will clearly influence the kind of spinors involved.}.
The   Dirac algebra is naturally decomposed in two subsets: the components along the worldsheet and those
orthogonal to the worldsheet, which in the ten-dimensional case means
\begin{eqnarray}
\label{gammas}
\rho_{\alpha}&=&t_{A \alpha} \Gamma^{A}\,,\qquad \qquad \qquad\qquad~~~   {\alpha=1,2}\,,\\
\rho_{\underline{i}}&=&N_{\underline{i}}^{A}\Gamma^{A} \,,\qquad \qquad \qquad~~~~~~~~~{\underline{i}=1,\dots,8}\,.
\end{eqnarray}
As used earlier in~\cite{Kavalov:1986nx,Sedrakian:1986np,Langouche:1987mw,Langouche:1987my,Langouche:1987mx} and made explicit in this context in~\cite{Drukker:2000ep},
since a two-dimensional Clifford algebra holds by construction for the  $\{\rho_\alpha,\rho_\beta\}=2\,\gamma_{\alpha\beta}$, it is always possible
 to find a \emph{local} $\grSO(1,9)$ rotation $S$   that transforms $\rho_i$ into two-dimensional Dirac matrices contracted
with zweibein
\be
\begin{split}\label{gammasrot}
\rho_{\alpha}&=S\Gamma^{a} S^{-1} e_{a\alpha} \ \ \ \  {a=1,2}\,,\\
\rho_{\underline{i}}&=S\Gamma^{\underline{i}+2}S^{-1} \ \ \ \ \ \ \underline{i}=1,...,8 ,
\end{split}
\ee 
where  
 \begin{equation}\label{gammasfin}
\Gamma^1=i\,\tau_2\otimes \mathbf{1}_{16}\,,\qquad\qquad \Gamma^2=\tau_1\otimes \mathbf{1}_{16}\,,\qquad\qquad \Gamma^{\underline{i}}=\tau_3\otimes \Sigma^{\underline{i}} \,,\end{equation}
 $\tau_a$ are Pauli matrices and $\Sigma^{\underline{i}}$ are (16-dimensional) Dirac matrices in 8 Euclidean dimensions.
%
%
Defining now
\be
\Theta=S^{-1}\theta\  \ \ \mathrm{and}\ \ \  \ \hat\Omega_{\alpha}=S^{-1}\Omega_{\alpha}S+ S^{-1}\partial_{\alpha} S\,,
\ee
one ends  with the following rotated expression
\be\label{lagrferm2}
\tilde{L}_F^{kin}= 2i\sqrt{\g}\g^{\alpha\beta}\, 
\bar\Theta  \Gamma^{a} \, e_{a\alpha}\,(\partial_{\beta}+\hat \Omega_{\beta})\,\Theta.
\ee
We remark here that the present analysis is only valid at classical level: as we will see later the local rotation $S$ produces quantum mechanically a non-trivial Jacobian in the path-integral measure, whose contribution is crucial to recover the correct structure of the divergent terms. To evaluate $\hat\Omega_{\alpha}$, we begin expanding
\be
\Omega_{\beta}=S\hat\Omega_{\beta}S^{-1} -\partial_{\beta}S S^{-1}
\ee
 in the basis provided by the Dirac algebra \eqref{gammasrot} 
 \begin{eqnarray} 
\!\!\!\!\!\!\!\!\!\!\!
\frac{1}{4}\Omega_{\beta}^{AB}\Gamma_{AB}
\! =\! 
\frac{1}{4}(\hat \Omega_{\beta}^{ab}e_{a}^{\mu} e_{b}^{\nu}\rho_{\mu\nu}+2 \hat \Omega_{\beta}^{a \underline{i}} e_{a}^{\mu}\rho_{\mu \underline{i}}+
\hat\Omega^{\underline{i}\underline{j}}_{\beta}\rho_{\underline{i}\underline{j}})+\frac{1}{4}(\rho^{\alpha}\partial_{\beta}\rho_{\alpha}+\rho^{\underline{i}}\partial_\beta\rho_{\underline{i}}-\rho^{\alpha}\rho^{\mu} e^{a}_{\mu}\partial_\beta e_{a\alpha})\,.
\end{eqnarray}
The expression for $\partial_{\alpha} S S^{-1}$ in the second parenthesis above has been derived considering the derivatives of \eqref{gammasrot} with respect to 
worldsheet coordinates 
which in turn implies
\be
\rho^{\alpha}\partial_{\beta}\rho_{\alpha}+\rho^{\underline{i}}\partial_\beta\rho_{\underline{i}}
=-4 \partial_{\beta} S S^{-1}+\rho^{\alpha}\rho^{\mu} e^{a}_{\mu}\partial_\beta e_{a\alpha}.
\ee
%
Let us now formally expand all the contributions in the   same basis \eqref{gammas}
\begin{equation}
\Omega_{\beta}^{AB}\Gamma_{AB}=
\Omega_{\beta}^{AB} t_{A}^{\mu} t_{B}^{\nu}\rho_{\mu\nu}+
2\Omega_{\beta}^{AB} t_{A}^{\mu} N_{B}^{\underline{i}}\rho_{\mu \underline{i}}+
\Omega_{\beta}^{AB} N_{A}^{\underline{i}} N_{B}^{\underline{j}}\rho_{\underline{i}\underline{j}}
\ee
in which 
\begin{eqnarray}\nonumber
&&\!\!\!\!\!\!\!\!
\rho^{\alpha} \partial_{\beta}\rho_{\alpha}-\rho^{\alpha\mu} e^{a}_{\mu}\partial_\beta e_{a\alpha}=\rho^{\alpha \underline{i}}  N_{\underline{i}}^{A} K_{A\alpha\beta}-\rho^{\alpha\mu} e_{a\alpha} e_{b\mu}\omega^{ab}_{\beta}+
\rho^{\alpha\mu}\Omega^{AB}_{\beta} t_{A\alpha} t_{B\mu}
+\rho^{\alpha \underline{i}}\Omega^{AB}_{\beta} t_{A\alpha} N_{B \underline{i}}\\
&&\!\!\!\!\!\!\!\!
\rho^{\underline{i}} \partial_{\beta}\rho_{\underline{i}}
=\rho^{\alpha \underline{i}}  (N_{\underline{i}A} K^{A}_{\alpha\beta}+\Omega_{\beta}^{BA} t_{B\alpha}N_{\underline{i} A})+\rho^{\underline{i}\underline{j}} (N^{A}_{\underline{j}} D_{\beta} N_{\underline{i} A}+\Omega^{BA}_{\beta}N_{\underline{i} B} N_{A\underline{j}}).
\end{eqnarray}
Collecting the different results we obtain
\be
\hat\Omega_{\alpha}^{ab}e_{a}^{\rho} e_{b}^{\sigma}=
e_{a}^{\rho} e_{b}^{\sigma}\omega^{ab}_{\alpha}\,,\qquad\qquad
\hat\Omega_{\alpha}^{a \underline{i}} e_{a}^{\rho}
=-N^{i}_{A} K^{A\rho}_{\beta}\,,\qquad\qquad
\hat\Omega^{\underline{i}\underline{j}}_{\alpha}
=-A^{\underline{i}\underline{j}}_{\beta}\,,
\ee
and the fermion kinetic part of the rotated Lagrangian takes the form
\be\label{lagrfermkin}
\begin{split}
\tilde{L}_F^{kin}=&2i\,\sqrt{\g}\g^{\alpha\beta}\,
\bar\Theta \, \Gamma^{a} \, e_{a\alpha}\, \big(\partial_{\beta}+\textstyle{\frac{1}{4}}\Gamma_{bc}\omega^{bc}_{\beta}-
\frac{1}{4}N^{\underline{i}}_{A}K^{A b}_{\beta}\Gamma_{b \underline{i}}-\textstyle{\frac{1}{4}} A^{\underline{i}\underline{j}}_{\beta} \Gamma_{\underline{i}\underline{j}}\,\big)\Theta\\
\equiv&2i\,\sqrt{\g}\g^{\alpha\beta}\,
\bar\Theta \, \Gamma^{a} \, e_{a\alpha}\, \big(\,\partial_{\beta}+\textstyle{\frac{1}{4}}\Gamma_{bc}\omega^{bc}_{\beta}-\textstyle{\frac{1}{4}} A^{\underline{i}\underline{j}}_{\beta} \Gamma_{\underline{i}\underline{j}}\,\big)\,\Theta \,.
\end{split}
\ee
 Namely,  in the rotated basis \eqref{gammasrot}-\eqref{gammasfin}  the GS
kinetic operator \eqref{lagrfermkinRT} results in a 
standard two-dimensional Dirac fermion action
on a curved two-dimensional background with geometry defined by the induced metric~\cite{Drukker:2000ep}. The spinor covariant derivative can be written as a two-dimensional, ordinary, spinor covariant derivative plus two additional  terms as in the first line above. The first extra-term is proportional  to the extrinsic curvature, it mixes tangential and normal components and  drops out naturally in the fermionic action -- hence the second line in \eqref{lagrfermkin} -- once contracted with $\g^{\alpha\beta} \Gamma^{a} \, e_{a\beta}$ and $\varepsilon^{\alpha\beta} \Gamma^{a} \, e_{a\beta}$. In fact, using the equations of motion \eqref{meanvanish} and because of the symmetry of $K^{m}_{\alpha\beta}$ in $\alpha,\beta$, it holds
\be
\begin{split}
\!\!\!\!\!\!&K^{A ab}\Gamma_{a}\Gamma_{b\underline{i}}=K^{Aab}(\Gamma_{ab\underline{i}}+\delta_{ab}\Gamma_{\underline{i}})=0\,,\\
\!\!\!\!\!\!& \varepsilon ^{\alpha \beta } e^a{}_{\alpha } K_{\beta}^{Ab} \Gamma _a\Gamma _{b\underline{i}}
=\varepsilon ^{\alpha \beta } e^a{}_{\alpha } K_{\beta }^{Ab} (i \varepsilon _{ab} \tau_3+\delta_{ab})\Gamma_{\underline{i}}= \big(i \tau_3  \det (e) e_b^{\beta }  K_{\beta }^{ Ab} +\varepsilon ^{\alpha \beta }  K_{\alpha \beta }^A\big)\Gamma_{\underline{i}}=0\,\,.
\end{split}
\ee
The second additional term consists of an extra, normal bundle two-dimensional gauge connection $ A^{\underline{i}\underline{j}}_{\alpha}$, with respect to which the 16 two-dimensional spinors making up the  32-component MW spinor $\Theta$  transform  in the spinorial representation of  $\grSO(8)$.  This  interacting term  vanishes (\emph{i.e.} a choice of the $N^{\underline{i}}$ exists such that $ A^{\underline{i}\underline{j}}_{\beta}=0$) when the  field-strength   associated to the normal connection vanishes, see discussion around \eqref{Fij}. 
As mentioned there, this is always the case, for example,  for embeddings of the string worldsheet in $\ads_3$,  where indeed the normal direction is just one and the normal bundle is then trivial. 
For a more general embedding  extending in both $\ads_5\times \sphere^5$ subspaces, the presence or not of this interaction term has to be checked case by case. This is what we do in Section \ref{sec:applications}, where in considering several examples also comment on this aspect. 

\subsection{Fermionic ``mass'' matrix: the flux term}
\label{sec:fermions_flux_term}

We now analyze the flux term \eqref{fluxstar} in the fermionic Lagrangian, \eqref{fermionGS}-\eqref{def_covariant_der} which after the $\kappa$-symmetry gauge-fixing $\theta^1=\theta^2\equiv\theta$ reads
\be\label{lagrfermflux2}
L_{F}^{flux}=-\varepsilon^{\alpha\beta}\,\bar\theta \,\rho_\alpha\,\Gamma _* \rho_\beta\,\theta\,\,, \qquad~~~ \Gamma_* =i \Gamma_{01234}\,.
\ee 
In order to understand the geometrical meaning of the terms in \eqref{lagrfermflux2}, we again decompose the Gamma matrices in the orthonormal basis formed by the tangent and transverse vectors. Remembering that $\Gamma_*$ contains only ten-dimensional flat indices belonging to $\ads_5$, we have
\begin{eqnarray}
\varepsilon ^{\alpha\beta} \rho _\alpha \Gamma _* \rho _\beta &=&
\varepsilon ^{\alpha\beta}\Gamma_*  \left(\Gamma _A \hat{t}_\alpha^A-\Gamma_A \bar{t}_\alpha^A\right)\, \rho _\beta\\\nonumber
&=&\varepsilon ^{\alpha\beta} \Gamma _*  \left(\Gamma_B (\g^{\lambda\delta}  t_\lambda^A t_\delta^B+N_{\underline{i}}^A N_{\underline{i}}^B)\,\hat{t}_{\alpha\,A}
-\Gamma_B( \gamma^{\lambda\delta} t_\lambda^A t_\delta^B+N_{\underline{i}}^A N_{\underline{i}}^B )\bar{t}_{\alpha\,A}
\right)\, \rho _\beta\\\nonumber
&=&\varepsilon^{\alpha\beta} \Gamma _*  \left(\gamma^{\lambda\delta}  \rho _\lambda\rho_\beta \,[(\hat{t}_\delta\cdot\hat{t}_\alpha)-(\bar{t}_\delta\cdot\bar{t}_\alpha)]-2\rho _r\rho_\beta (\bar{N}_r\cdot\bar{t}_\alpha)\right)\,,~~~r=7,8\,
\end{eqnarray}
where we used the completeness relation $\eta^{AB}=\gamma^{\lambda\delta} t_\lambda^A t_\delta^B+N_{\underline{i}}^A N_{\underline{i}}^B$,  the fact that $N_{\underline{i}}^A \bar{t}_\alpha^A$, $N_{\underline{i}}^A \hat{t}_\alpha^A$ are non vanishing only for $\underline{i}=r\equiv7,8$ and that  $\hat N_r^A \hat{t}_\alpha^A=-\bar N_r^A \bar{t}_\alpha^A, \,r=7,8$~\footnote{Here we choose a basis for the  normal directions $\underline{i}=1,\dots,8$ which matches the one used in the bosonic analysis. Namely, fluctuations  \underline{i}=7,8 are  along two orthogonal directions with non-vanishing extrinsic curvature, and for the remaining $i=1,\dots, 6$ it holds \eqref{orthogonality}, with  three of them in $\ads_5$ and three in $\sphere^5$.}.
Defining the antisymmetric product of the Gamma matrices projected onto the worldsheet as 
\be\label{tau3}
\rho_3\equiv \frac{1}{2\sqrt{\gamma}} \varepsilon^{\alpha\beta}\rho_{\alpha\beta}\,,\qquad\qquad
\rho^{\alpha\beta}=- \frac{1}{\sqrt{\gamma}}\, \varepsilon^{\alpha\beta}\,\rho_{3}\,,
\ee
one can rearrange the flux term in the following way
\begin{eqnarray}\label{eqdimezzo}
\varepsilon ^{\alpha\beta} \rho _\alpha \Gamma _* \rho _\beta &=&
\varepsilon^{\alpha\beta} \Gamma _*  \left[\gamma^{\lambda\delta} ( \sqrt{\g}\, \varepsilon_{\lambda\beta}\,\rho _3 +\g_{\lambda\beta})  \,[(\hat{t}_\delta\cdot\hat{t}_\alpha)-(\bar{t}_\delta\cdot\bar{t}_\alpha)]-2\rho _r\rho_\beta \left(\bar{N}^r\cdot \bar{t}_\alpha\right)\right]\\\nonumber
&=& \Gamma _*  \left[ \sqrt{\g} \,\rho _3 \,\g^{\alpha\delta}   \,[(\hat{t}_\delta\cdot\hat{t}_\alpha)-(\bar{t}_\delta\cdot\bar{t}_\alpha)]-2\, \varepsilon ^{\alpha\beta}\,\rho _{r\beta}  \left(\bar{N}^r\cdot \bar{t}_\alpha\right)\right]\\\nonumber
&=& \Gamma _*  \left[  \sqrt{\g} \,\rho _3 \,( m^{2}_{AdS_{5}}+m^{2}_{S^{5}})+2\,\varepsilon ^{\alpha\beta}\,\rho _{r\beta}\, 
\nabla_{\lambda} {K^{r}_{\alpha}}^\lambda\right]\,,
\end{eqnarray}
where we used that $\varepsilon ^{\alpha\beta} \left({t}_\alpha\cdot{t}_\beta\right)=0$,  the definition of masses \eqref{massesads5s5}, and the Gauss-Codazzi equation \eqref{codazzi}~\footnote{
More precisely, from \eqref{codazzi}, it immediately follows
\be\nonumber
\mathcal{D}_{\alpha} K^{i}_{\beta\gamma}-\mathcal{D}_{\beta} K^{i}_{\alpha\gamma}
=
-(\bar t_{\alpha}\cdot t_{\gamma})(\bar t_{\beta}\cdot N^{i})
+(\bar t_{\alpha}\cdot N^{i})(\bar t_{\beta}\cdot t_{\gamma})
+(\hat t_{\alpha}\cdot t_{\gamma})(\hat t_{\beta}\cdot N^{i})
-(\hat t_{\alpha}\cdot N^{i})(\hat t_{\beta}\cdot t_{\gamma}).
\ee
Since $(\bar t_{\beta}\cdot N_{i})=-(\hat t_{\beta} \cdot N_{i})$, we find
\begin{align}\nonumber 
\mathcal{D}_{\alpha} K^{i}_{\beta\gamma}-\mathcal{D}_{\beta} K^{i}_{\alpha\gamma}=&
[(\bar t_{\beta}\cdot t_{\gamma})+(\hat t_{\beta}\cdot t_{\gamma})](\bar t_{\alpha}\cdot N^{i})
-
[(\bar t_{\alpha} \cdot t_{\gamma})+(\hat t_{\alpha}\cdot t_{\gamma})](\bar t_{\beta}\cdot N^{i})
=\gamma_{\beta\gamma} (\bar t_{\alpha}\cdot N^{i})
-\gamma_{\alpha\gamma} (\bar t_{\beta}\cdot N^{i})~.
\end{align}
Tracing on $\alpha$ and $\gamma$ and using the equations of motion \eqref{meanvanish}, one obtains
 \be\nonumber
 \mathcal{D}_{\alpha} K^{i~\alpha}_{\beta}=  -(\bar t_{\beta}\cdot N^{i})= (\hat t_{\beta}\cdot N^{i})~.
 \ee
}. 
Hence, after the rotation \eqref{gammasrot} and in terms of the spinors $\Theta=S^{-1}\theta$, the flux part of the fermionic Lagrangian takes the form 
\begin{eqnarray}\label{L_ferm_flux}
\!\!\!\!
\tilde{L}_{F}^{flux}=\bar\Theta \,\tilde{\Gamma}_*
\Big[ \sqrt{\g} \,\tau _3 \,( m^{2}_{AdS_{5}}+m^{2}_{S^{5}})+2\,\varepsilon ^{\alpha\beta}e^a_{\beta}\,\Gamma_{ar}\, 
\nabla_{\gamma} {K^{r}_{\alpha}}^\gamma\Big] \,\Theta\,\,,~~r=7,8\,,
\end{eqnarray}
where $\tau_3 = S^{-1}\rho_3\,S$, and $\tilde{\Gamma}_*=S^{-1}{\Gamma}_* S$. In general, the rotated $\Gamma_*$ is written as
\begin{flalign}\label{gammastartilde_def}
\tilde{\Gamma}_*  &= S^{-1}{\Gamma}_* S=
i S^{-1} (\hat{\epsilon}^{ABCDE}\Gamma_{A}\dots\Gamma_{E} )S
\\ \nn
&=i \hat{\epsilon}^{ABCDE}(\hat{t}_{A}^{a}\Gamma_{a}+\hat{N}_{A}^{i}\Gamma_{i}+N_{A}^{r}\Gamma_{r} )\dots (\hat{t}_{E}^{e}\Gamma_{e}+\hat{N}_{E}^{j}\Gamma_{j}+N_{E}^{s}\Gamma_{s} ) \,,
\end{flalign}
(where the hat in $\hat{\epsilon}^{ABCDE}$ is to signal that  the $A,B,C,D,E$ take values $0,1,2,3,4$, as clear from \eqref{lagrfermflux2}, $i, j=1\,, \dots\,, 6$, $r,s=7,8$) and it can be expanded in the same basis. 
A clever choice among the basis vectors made possible by the string motion, drastically simplifies the above expression.  Due to the self-duality of RR flux, we have written the flux contribution only in terms of $\Gamma_*$, and thus we can restrict our discussion and attention to the AdS projection. This is the basis we have in mind here with the aim to simplify the product \eqref{gammastartilde_def}.

In the more general case discussed in Section \ref{normalbundle}, in order to compute \eqref{gammastartilde_def}, we can use the basis  build up by the two tangent vectors  projected onto $\ads_5$ $\hat t_\alpha$ ($\alpha=1,2$) and the three transverse vectors $\hat N_i$ ($i=1,2,3$) which entirely lie in $\ads$. This set forms an orthonormal basis spanning $\ads$. However, there might be further contributions also from the remaining transverse directions $N_r$ ($r=7,8$) and for this we need to project them in the above basis, that is 
\be\label{n_expansion}
\hat N^r_A =\xi^r_a \, \hat t^a_A\,~\,,\qquad\qquad\,\,\,\xi^r_a= ( \hat N^r\,\cdot \hat t_\alpha )\hat g^{\alpha\beta} e_{\beta a}\,  \qquad r=7,8\,,
\ee
 where $\hat g^{\alpha\beta}$ is the inverse matrix obtained from worldsheet metric projected in the sub-space $\ads$, {\it i.e.} $\hat g_{\alpha\beta}\equiv (\hat t_\alpha \cdot \hat t_\beta)$, $\hat g^{\alpha\beta}\,\hat g_{\beta\delta}= \delta^a_\delta$. Clearly, an analogous expression to \eqref{n_expansion} can be written for the projected vectors onto the compact sub-space, with the obvious replacement from hatted to bar vectors. By means of the relation \eqref{n_expansion} we can massage the product \eqref{gammastartilde_def} to the final expression
\begin{flalign}\label{gammastartilde_better}
\tilde{\Gamma}_* =
i \hat{\epsilon}^{ABCDE}
\hat{N}_{C}^{i}\hat{N}_{D}^{j}\hat{N}_{E}^{k} \Gamma_{ijk}
\left[ \hat{t}_{A}^{a}\hat{t}_{B}^{b}\, \left( \Gamma_{ab}+\xi^{[r}_a  \xi^{s]}_b\Gamma_{rs}\right)+
\half \xi^r_d \left(\hat t^c_A \hat t^d_B-\hat t^c_B \hat t^d_A\right)\Gamma_{cr}\right]. ~~~~
\end{flalign}
In writing the above expression we are assuming that the inner product \eqref{n_expansion} is not degenerate (as well as the projected worldsheet metric $\hat g_{\alpha\beta}$). 
However, whenever (at least) one of the tangent vectors has a vanishing projection in $\sphere^5$, the number of transverse directions completely lying in the compact sub-space increases (of at least one), since now we need (at least) another transverse direction to span $\sphere^5$. By a simple counting of degrees of freedom, this decreases (at least) by one the remaining $\hat N$. This implies that one of the transverse direction $N^r$ ($r=7,8$) can become linearly independent and lie completely in $\ads_5$, that is the inner product \eqref{n_expansion} in one of the directions $r=7,8$ could be degenerate. This happens for example in the case of the spinning string motion \cite{Frolov:2002av} discussed in Section \ref{app:spinning}. 
In the limiting case when the motion is completely embedded in $\ads_3\subset \ads_5$, the tangent vectors belong only to $\ads$, and both the directions $N^r$ can be chosen such that one will completely lie in $\ads$ and the other one in $\sphere^5$. Then, we can pick a basis given by $t_\alpha$, $\alpha=1,2$ (since now $\hat t_\alpha=t_\alpha$) and three transverse directions $\hat N_i$ (i=1,2), $N_r$ with $r$ either 7 or 8. The remaining vectors belong only to the sphere. In this case there is only one term contributing to the expression \eqref{gammastartilde_def}, namely the first one in \eqref{gammastartilde_better}. We illustrate two examples where this takes place: the folded string \cite{Frolov:2002av} at the end of  Section  \ref{app:spinning}, and the string configuration dual to the circular Wilson loop operator \cite{Drukker:2000ep} at the end of Section \ref{app:latitude}.  If the minimal surface is completely embedded in $\sphere^5$ then this means that  only the transverse directions will contribute to the product \eqref{gammastartilde_def}, that is 
\be
\tilde{\Gamma}_* =S^{-1}{\Gamma}_* S=\hat{\epsilon}^{ABCDE} \hat N^r_A \hat N^s_B \hat{N}_{C}^{i}\hat{N}_{D}^{j}\hat{N}_{E}^{k} \Gamma_{rsijk}\,. 
\ee


Finally, we want to stress that the arguments explained above work in a completely symmetric manner by swapping the two sub-spaces $\ads$ and $\sphere$.
Due to the self-duality of the flux term, we could have chosen $\Gamma_*=i \Gamma_{56789}$ and then write $\tilde{\Gamma}_*= i \bar \epsilon_{ABCDE} \bar N^r_A \bar N^s_B \bar{N}_{C}^{i}\bar{N}_{D}^{j}\bar{N}_{E}^{k} \Gamma_{rsijk}$ for an $\ads_3$ motion, keeping also in mind an overall sign originated in the manipulations in equation \eqref{eqdimezzo}. However, we prefer to keep on working with the flux defined in \eqref{gammastartilde_def} for reader's convenience.

\noindent
 

\subsection{Comparison with bosonic masses and quantum divergences}

In this section we work out a sum rule for the fermion ``mass matrix'' in analogy to the bosonic one \eqref{sumrulebos}. In this last case the trace of the mass matrix ${\cal M}$ appears naturally in the Seeley-De Witt coefficient $a^{(2)}_B$, controlling the logarithmic divergence of the bosonic fluctuactions \cite{Gilkey} 
\be\label{seeley_coeffbos}
a^{(2)}_B=\mathrm{Tr}(\frac{1}{6}\,R^{(2)}+{\cal M}).
\ee
A natural candidate for the traced fermionic mass matrix is therefore the companion term appearing in the fermionic Seeley-De Witt coefficient $a^{(2)}_F$, driving the logarithmic fermionic divergences as well. In the most general case, for a self-adjoint Dirac operator $i \rho^\alpha\mathcal{D}_\alpha-\mathcal{M}_F$  this is given by \cite{Gilkey}%
%
%
\be\label{seeley_coeff}
a^{(2)}_F=\mathrm{Tr}(\mathbb{I})\frac{R^{(2)}}{12}-\frac{1}{2}\mathrm{Tr}\left(\rho^\alpha \mathcal M_F \rho_\alpha \mathcal M_F\right)\,,
\ee
and we are at the moment interested in the second term above.
In our case, the fermionic mass matrix is 
\be\label{def_mf}
\mathcal M_F = {1\over 2 \sqrt{\gamma}} \ve^{\alpha\beta} \rho_\alpha\Gamma_*\rho_\beta\,,
\ee
where we re-wrote  the fermionic Lagrangian as
 \be
\mathcal L_{2F}= 2 i \sqrt{\gamma}\, \bar\theta \,\left(\gamma^{\alpha \beta} \rho_\alpha\mathcal{D}_\beta +{i\over 2 \sqrt{\gamma}} \epsilon^{\alpha\beta} \rho_\alpha\Gamma_*\rho_\beta\right)\theta
:= 2 i \sqrt{\gamma}\, \bar\theta D_F \theta\,.
\ee
We recall that our fermions are worldsheet scalars and the square root of the determinant of the induced metric is the correct normalization in the fermionic norms \cite{Drukker:2000ep}. It is important to stress that here $\gamma$ is the absolute value of the determinant. 
When $\rho_\alpha$ commutes with $\mathcal M_F$ the invariant \eqref{seeley_coeff} reduces to the more familiar $\mathrm{Tr}\left(\mathcal M_F^2\right)$, analysed in \cite{Drukker:2000ep} or in \cite{Drukker:2011za} for example.  This is the case for the folded spinning solution reviewed in Section \ref{app:spinning}. However, for a general string solution, like the latitude configuration, the two matrices do not commute and \eqref{seeley_coeff} is an invariant, which leads to the sum rule as we explain below. 

Notice that due to the ciclicity of the trace we can compute \eqref{seeley_coeff} before performing any local $\grSO(1,9)$ rotations, and actually in this case it turns out a convenient strategy. 
Let us start by rewriting \eqref{seeley_coeff} as
\begin{flalign}
\mathrm{Tr}\left(\rho^\alpha \mathcal M_F \rho_\alpha \mathcal M_F\right) &=
\mathrm{Tr}\left({\gamma^{\alpha\beta}\over 4\gamma} \ve^{\mu\nu} \ve^{\lambda\tau} 
\rho_\alpha\rho_\mu \Gamma_* \rho_\nu \rho_\beta \rho_\lambda \Gamma_* \rho_\tau \right)
\\ \nn &=
\mathrm{Tr}\left({\gamma^{\alpha\beta}\over 4\gamma} \ve^{\mu\nu} \ve^{\lambda\tau} \rho_\alpha\rho_\mu \Gamma_* 
\left(\gamma_{\nu\beta}\rho_\lambda+\gamma_{\beta\lambda} \rho_\nu-\gamma_{\nu\lambda}\rho_\beta\right) \Gamma_* \rho_\tau\right)
\\ \nn
&\equiv I_1+I_2+I_3\,. 
\end{flalign}
and compute the three terms above separately. 
The first contribution $I_1$ gives
\begin{flalign}
I_1 &={1\over 4\gamma} \mathrm{Tr}\left(\ve^{\mu\nu} \ve^{\lambda\tau}\rho_\nu \rho_\mu \Gamma_*\rho_\lambda \Gamma_* \rho_\tau \right)
\\ \nn
&=
-{1\over 4\gamma}  \mathrm{Tr}\left(\ve^{\mu\nu}\rho_\mu \rho_\nu\, \Gamma_* \Gamma_*
 \left( \sqrt{\gamma} \left(m_{AdS_5}^2+m_{S^5}^2\right) \rho_3+2 \ve^{\alpha\beta} \left(\hat t_\alpha \cdot N^r\right) \rho_r \rho_\beta\right)\right)\,,
\end{flalign}
where we have used that 
\be\label{aux1}
\ve^{\alpha\beta}\rho_\alpha\Gamma_*\rho_\beta= 
\Gamma_*\left\{
\sqrt{\gamma}\left( m^2_{AdS_5}+m^2_{S^5}\right)\rho_3 +2\, \ve^{\alpha\beta} \left(\hat t_\alpha\cdot N^r\right) \rho_r \rho_\beta\right\}\,.
\ee
computed in Section \ref{sec:fermions_flux_term}. 
By recalling that $\rho_3={1\over 2\sqrt{\gamma}}\ve^{\alpha\beta}\rho_{\alpha\beta}\,$, and $\rho^{\alpha\beta}= -{\ve^{\alpha\beta}\over \sqrt{\gamma}} \rho_3$, we can compute the following commutation relation
\be
\rho_3 \rho_\beta={ \ve^{\mu\nu}\over \sqrt{\gamma}} \gamma_{\nu\beta}\rho_\mu\,,
\ee
which in $I_1$ leads to the result
\begin{flalign}\label{term_I1}
I_1 &= -\half \left(m_{AdS_5}^2+m_{S^5}^2\right)\mathrm{Tr}\left(\rho_3^2\right)\,.
\end{flalign}
The second term $I_2$ can be manipulated in a similar manner and we obtain
\begin{flalign}\label{term_I2}
I_2 &={1\over 4\gamma} \mathrm{Tr}
\left( \gamma^{\alpha\beta}\ve^{\mu\nu} \ve^{\lambda\tau} \rho_\alpha\rho_\mu \Gamma_* 
\gamma_{\beta\lambda} \rho_\nu \Gamma_* \rho_\tau \right)=
-\half \left( m^2_{AdS_5}+m^2_{S^5}\right)\mathrm{Tr}(\mathbb{I})\,.
\end{flalign}
Finally let us look at the last term $I_3$
\begin{flalign}\label{term_I3}
I_3 &= 
-{1\over 4 \gamma} \mathrm{Tr}\left(\gamma^{\alpha\beta}\ve^{\mu\nu}\ve^{\lambda\tau} \gamma_{\nu\lambda}\rho_\alpha\rho_\mu\Gamma_*\rho_\beta\Gamma_*\rho_\tau\right) \\ \nn
&=
-{1\over 4}  \mathrm{Tr}\left(\gamma^{\alpha\beta}\Gamma_*\rho_\alpha\Gamma_*\rho_\beta\right)
+{1\over 4} \mathrm{Tr}\left( \gamma^{\alpha\beta}\gamma^{\mu\nu}\rho_{\mu\alpha}\Gamma_* \rho_\beta\Gamma_*\rho_\nu \right)\,.
\end{flalign}
In order to evaluate $I_3$ we need also the expression for the symmetrized product of Gamma matrices, that is
\be\label{aux2}
\gamma^{\alpha\beta} \rho_\alpha\Gamma_* \rho_\beta= \Gamma_*\left[ \left(m^2_{AdS_5}+m^2_{S^5}\right) +2 \left(\hat t_\alpha\cdot N_r\right) \rho^r \rho^\alpha\right]\,,
\ee
which together with \eqref{aux1} leads to
\begin{flalign}\label{term_I3_final}
I_3 &=-{1\over 4}\left( m^2_{AdS_5}+m^2_{S^5}\right)\, \mathrm{Tr}(\mathbb{I}-\rho_3^2)\,. 
\end{flalign}
Hence, collecting the three contributions \eqref{term_I1}, \eqref{term_I2}, and \eqref{term_I3_final}, and using $\rho_3^2= \mathbb{I}$, we obtain
\be\label{trm_fermions}
\mathrm{Tr}\left(\rho^\alpha \mathcal M_F \rho_\alpha \mathcal M_F\right) =  -\left( m^2_{AdS_5}+m^2_{S^5}\right)\mathrm{Tr}(\mathbb{I})= \left( ^{(2)}\!\! \,R+\mathrm{Tr}(K^2)\right)\mathrm{Tr}(\mathbb{I})\,
\ee
by means of the expressions \eqref{ms5mads5}. Armed with  \eqref{trm_fermions} we can compare the bosonic and fermionic contribution to the total logarithmic divergences, by computing explicitly  $a^{(2)}_B+a^{(2)}_F$. The part depending on the traced mass matrices is
\be
\label{sum_rule_all}
-\half \mathrm{Tr}\left(\rho^\alpha \mathcal M_F \rho_\alpha \mathcal M_F\right)+\mathrm{Tr}\left(\mathcal M\right)=
-4\left( ^{(2)}\!\! \,R +\mathrm{Tr}(K^2)\right) + 3 \,^{(2)}\!\! \,R + 4 \mathrm{Tr}(K^2)= -\, ^{(2)}\!\! \,R\,,
\ee
We have used $\mathrm{Tr}(\mathbb{I})=8$ since these are eight $2d$ fermionic modes now, as reviewed at the beginning of Section \ref{sec:fermions_kin}.  The contribution of the pure intrinsic curvature terms is instead more subtle \cite{Drukker:2000ep}. The use of the naive Seeley-De Witt fermionic coefficient would lead to a wrong result
\be\label{wrong}
8\frac{1}{6}\,R^{(2)}+8\frac{1}{12}\,R^{(2)}=2R^{(2)},
\ee
that combined with \eqref{sum_rule_all} would not produce the correct coefficient $3R^{(2)}$  \cite{Drukker:2000ep}. The reason of the apparent disagreement is well known \cite{Langouche:1987mw,Langouche:1987my,Langouche:1987mx, Drukker:2000ep}: the local $\grSO(1,9)$ rotation $S$  that transforms $\rho_i$ into two-dimensional Dirac matrices contracted
with zweibein gives rise to a non-trivial Jacobian in the path-integral measure, that contributes additionally to the logarithmic divergence. The net effect of this ``anomalous rotation'' is to change the coefficient of the conformal anomaly of the relevant two-dimensional Dirac fermions by a factor 4. In our case it amounts to modify the fermionic contribution to \eqref{wrong} by a factor 4 and therefore
 \be\label{right}
8\frac{1}{6}\,R^{(2)}+8\frac{1}{3}\,R^{(2)}=4R^{(2)},
\ee
recovering, in combination with \eqref{sum_rule_all}, the result of \cite{Drukker:2000ep}.


\section{Applications}
\label{sec:applications}

In this Section we work out a few relevant situations in which the general structures previously derived are exemplified. First we discuss the spinning string~\cite{Gubser:2002tv,Frolov:2002av}, recovering the known spectrum of the fluctuations. Then we apply our analysis to the minimal surface associated to the 1/4 BPS Wilson loop operator~\cite{Drukker:2005cu, Drukker:2006ga, Drukker:2007qr}, obtaining the bosonic and fermionic sectors and the related mass matrices.

\subsection{Spinning strings}
\label{app:spinning}

The first test of our analysis is the spinning string solution, generalization of~\cite{Gubser:2002tv} and~\cite{Berenstein:2002jq}, whose semiclassical quantization was worked out in~\cite{Frolov:2002av}~\footnote{See \cite{Forini:2012bb} for the fermionic mass matrix of the string rotating  in $\ads_5\times \sphere^1$.}.
We choose the following coordinates for the target space metric
\be
\begin{split}
\left.ds^{2}\right|_{AdS^{5}} & =-\cosh^{2}\rho\, d{t}^{2}+d\rho^{2}+\sinh^{2}\rho\left(d\beta_1^{2}+\cos^{2}\beta_1 d\beta_2^{2}+\cos^2 \beta_1 \cos^2\beta_2 d\phi^{2}\right)\,,
\\
\left.ds^{2}\right|_{S^{5}} & =d\psi_1^{2}+\cos^{2}\psi_1 \left[d\psi_2^{2}+\cos^{2}\psi_2\left(d\psi_3^{2}+\cos^2\psi_3d\psi_4^2+\cos^2\psi_3\,\cos^2\psi_4d\varphi^2\right)\right]\,.
\end{split}
\ee
The motion of the spinning string in $\ads_5 \times \sphere^5$ is described by the following ansatz for the embedding coordinates \cite{Frolov:2002av}
\be\label{spinning_ansatz}
t=\kappa \tau\,, \quad \rho=\rho(\sigma)\,, \quad \beta_i=0 \,~ (i=1,2)\,, \quad \phi=\omega\tau \,, \quad \psi_i=0\,~ (i=1, \dots , 4)\,, \quad \varphi=\nu \tau\,, 
\ee
where $(\tau\,, \sigma)$ are the worldsheet coordinates, and $\rho(\sigma)$ has to satisfy the equation of motion and  Virasoro constraints
\be
\rho''= \left(\kappa^2-\omega^2\right)\cosh\rho  \sinh\rho \,, \qquad
{\rho' }^2 = \kappa^2\cosh^2\rho  -\omega^2 \sinh^2\rho  -\nu^2\,. 
\ee
The solution $\rho(\sigma)$ can be found explicitly and it reads in terms of Jacobi elliptic functions, since we will not need it here we refer the reader to \cite{Frolov:2002av} for further details.
The parameters $\kappa, \omega,$ and $\nu$  are chosen to be positive, and moreover the finite energy condition as well as the request of having a real solution impose certain relations on the parameters, that is
%
%
\be
\kappa > \nu\,, \qquad \omega> \nu\,, \qquad \coth^2\rho  \ge {\kappa^2-\nu^2\over \omega^2-\nu^2} \,.
\ee

In order to compute the quadratic fluctuation Lagrangian, we need to construct an orthonormal basis as in Section \ref{sec:mass_matrix_sum_rules}. The tangent vectors are given by
%
\be
t^A_\tau = \left( \kappa \cos\rho\,, 0\,, 0\,, 0\,, \omega \sinh\rho \,, 0\,, 0\,, 0\,,0\,, \nu\right)\,,\qquad 
t^A_\sigma =\left(0\,, \rho'\,,0\,, 0\,, 0\,,0\,,0\,, 0\,, 0\,,0 \right)\,.
\ee
Then, we choose the 8 orthogonal vectors as follows
\begin{flalign}\label{spinning_basis}
\hat N^{A}_1 &= \left(0\,, 0\,,1\,,  0\,, 0\,, 0\,,0\,,  0\,, 0\,, 0\,\right)\,, \qquad
\hat N^A_2=  \left( 0\,, 0\,, 0\,,1\,,  0\,, 0\,, 0\,,0\,,  0\,, 0 \,\right)\,,\\ \nn
\bar N^A_1 &=\left( 0\,, 0\,, 0\,,0\,,  0\,, 1\,,  0\,, 0\,, 0\,,0\right)\,, \qquad 
\bar N^A_2 = \left( 0\,, 0\,, 0\,,0\,,  0\,, 0\,, 1\,,  0\,, 0\,, 0\right)\,,\\ \nn
\bar N^A_3 &= \left(  0\,, 0\,, 0\,,0\,,  0\,, 0\,,0\,, 1\,,   0\,, 0\right)\,,\qquad
\bar N^A_4 = \left( 0\,, 0\,, 0\,,0\,,  0\,, 0\,,0\,, 0\,, 1\,,  0\right)\, \\ \nn
N^A_7 &=\textstyle{ {1\over \sqrt{{\rho' }^2+\nu^2}}}\left( \omega \sinh\rho\,, 0\,, 0\,, 0\,, \kappa \cosh\rho\,, 0\,, 0\,, 0\,,0\,,  0\right)\,,\\ \nn
N^A_8 &=\textstyle{ {1\over  \rho '  \sqrt{\nu ^2+\rho '^2}} }
   \left(\kappa  \nu  \cosh\rho \,, 0\,, 0\,, 0\,, \nu \, \omega  \sinh \rho\,, 0\,,0\,,0\,,0\,,\nu ^2+\rho '^2\right)\,.
\end{flalign}
Finally, the induced worldsheet metric is given by
\be
\label{induced_ws_spinning}
\gamma_{\alpha\beta}=(t_\alpha\cdot t_\beta)= {\rho'}^2\diag\left(-1\,, 1\right)\,.
\ee
Once the basis vectors have been chosen, then the computation of the quadratic fluctuation Lagrangian is simply reduced to a mere application of expressions \eqref{L1}, \eqref{lagrfermkin}, \eqref{L_ferm_flux}.

\paragraph{Bosonic Lagrangian}
Let us proceed to the construction of the bosonic transverse Lagrangian \eqref{L1}.
The worldsheet curvature $^{(2)}\!\!\, R $ and the trace of the extrinsic curvature $\mathrm{Tr}(K^2)$ turn out to be
\be
\label{R2&K2_spinning}
^{(2)}\!\!\, R =\frac{2 \left(\kappa ^2-\nu ^2\right) \left(\omega ^2-\nu ^2\right)}{\rho '(\sigma )^4}-2 \,, \qquad 
\mathrm{Tr}(K^2)= 
-\frac{2 (\kappa^2 -\nu^2 ) ( \omega^2-\nu^2 )}{\rho '(\sigma )^4}-\frac{2 \nu
   ^2}{\rho '(\sigma )^2}\,.
\ee
The only non-zero components of the extrinsic curvature projected on the transverse directions,
{\it i.e.} $K^i_{\alpha\beta}\equiv N^i_A K^A_{\alpha\beta}$  
are
\be
K^7_{\tau\sigma} = K^7_{\sigma\tau} =\textstyle{ {\kappa\, \omega \rho'\over \sqrt{\nu^2+{\rho'}^2}}}\,, \qquad
K^8_{\tau\sigma} = K^8_{\sigma\tau} = -{\nu\rho''\over \sqrt{\nu^2+{\rho'}^2}}\,.
\ee
This implies that the mass matrix $\mathcal M_{ij}$ \eqref{L1} for $i, j=1\,, \dots\,, 6$ is diagonal and given by
\begin{flalign}
\mathcal M_{11} &=\mathcal M_{22}= - m^2_{\ads_5} =-\textstyle{\gamma^{\alpha\beta} \left(\hat t_\alpha\cdot \hat t_\beta\right) = -2-{\nu^2\over {\rho'}^2}}\,,  \\ \nn
\mathcal M_{33} &=\mathcal M_{44}=\mathcal M_{55} =\mathcal M_{66}= -m^2_{\sphere^5}=\textstyle{\gamma^{\alpha\beta} \left(\bar t_\alpha\cdot \bar t_\beta\right) =- {\nu^2\over {\rho'}^2}}\,.
\end{flalign}
Along the directions $7\,,8$ the extrinsic curvature contributes to the mass (in particular, the contribution to the off-diagonal matrix is entirely due  to the extrinsic curvature), and we obtain 
\begin{flalign}
\mathcal M_{77} &= -2 - {\nu^2\over {\rho'}^2} -{2\kappa^2\omega^2\over  {\rho'}^2 (\nu^2+{\rho'}^2)}\,,  \\ \nn 
\mathcal M_{88} &=-  {\nu^2\over {\rho'}^2}  +{2\kappa^2\omega^2\over  {\rho'}^2 (\nu^2+{\rho'}^2)}+{2(\kappa^2-\nu^2)(\nu^2-\omega^2)\over  {\rho'}^4}\,,\\ \nn
\mathcal M_{78} &= {2 \kappa \nu \omega \rho''\over  {\rho'}^3 (\nu^2+{\rho'}^2)}\,.
\end{flalign}
In order to complete the Lagrangian we construct the covariant derivative, for which we need the connection on the normal bundle $A^{ij}_\alpha$ \eqref{normalconnection}. The only non-trivial directions are along the transverse modes $7, 8$
\be\label{spinning_connection}
A_\sigma^{78}= -A_\sigma^{87}= - {\kappa\omega\nu\over \nu^2+{\rho'}^2}\,, \qquad A_\tau^{ij}=0\,, \qquad i, j=1, \dots, 8\,.
\ee
Hence, the bosonic Lagrangian agrees with \cite{Frolov:2002av}. 

\paragraph{Fermionic Lagrangian}

The construction of the fermionic Lagrangian proceeds in two steps: the kinetic part \eqref{lagrfermkin} and the flux term \eqref{L_ferm_flux}. 
For the kinetic term the only new ingredient we need is the worldsheet spin connection $\omega_{\alpha\, ab}$,
\be
\omega_{\tau\, 01}=-\omega_{\tau\, 10}= -{\rho''\over \rho'}\,, \qquad \omega_{\sigma\, a b}=0 \,. 
\ee
Hence, the kinetic term \eqref{lagrfermkin} is simply given by~\cite{Forini:2012bb}
\be\label{spinning_kin}
\mathcal L_{kin}= 2 i\,  \bar \Theta \left( \sqrt{\gamma}\gamma^{\alpha\beta} \Gamma_a e^a_\alpha \p_\beta+ \half \rho''\Gamma_1 +\half {\kappa\nu\omega\rho'\over (\nu^2+{\rho'}^2)} \Gamma_{189}\right)\Theta\,.
\ee
where we have used \eqref{spinning_connection} for the third term above, which is the connection-related part. Notice that, as normal bundle connection \eqref{spinning_connection} is \emph{flat}, {\it i.e.} $\p_\tau A_\sigma^{78}-\p_\sigma A^{78}_\tau+ [A_\tau\, A_\sigma]^{78}=0$, we can introduce a rotation which cancels such term - this is what done in \cite{Forini:2012bb}. 

In order to compute the flux term \eqref{L_ferm_flux} we cannot use the final expression \eqref{gammastartilde_better} of the expansion \eqref{gammastartilde_def}. This is due to the fact that for the spinning string one of the projected transverse vector, {\it i.e.} $\hat N_7$ in our convention, is orthogonal to $\hat t_\alpha$. A better choice to simplify the term $S^{-1} \Gamma_\star S$ is then the basis formed by the two tangent vectors $\hat t_\alpha$, the two transverse vectors $\hat N_1\,, \hat N_2$ and finally $\hat N_7$. The expansion \eqref{n_expansion} is still valid for $\hat N_8$. 
Thus, the only non zero terms which can contribute to $S^{-1}\Gamma_\star  S$ are 
\begin{flalign}\label{gammatilde_star_spinning2}
S^{-1}\Gamma_\star S &=
i \hat \epsilon_{ABCDE} \left( \hat t^A_\alpha \hat t^B_\beta e^\alpha_a e^\beta_b \hat N_i^C \hat N_j^D \hat N_r^E \Gamma_{ab} \Gamma_{ij} \Gamma_r 
+ \hat t^A_\alpha e^\alpha_a \hat N_i^B \hat N_j^C \hat N_r^D \hat N^E_s \Gamma_a \Gamma_{ij}\Gamma_{rs}\right)=\\ \nn
&=
i {\sqrt{\nu^2+{\rho'}^2} \over \rho'}\Gamma_{01238}- i{\nu\over \rho'}\Gamma_{12389} \,.
\end{flalign}
The remaining terms are straightforward to compute, simply from the inner products of the basis vectors, and we obtain
\begin{flalign}\nonumber
\mathcal L_{flux} &= {i \over 2} 
\bar\Theta \left( 2 (\nu^2+{\rho'}^2) \Gamma_{01}+2\nu \sqrt{\nu^2+{\rho'}^2 }\Gamma_{19}\right) \left(\textstyle{{\sqrt{\nu^2+{\rho'}^2} \over \rho'}}\Gamma_{01238}- \textstyle{{\nu\over \rho'}}\Gamma_{12389} \right) \Theta \\\label{spinning_flux}
&= i \rho' \sqrt{\nu^2+{\rho'}^2 }\, \bar\Theta\Gamma_{238}\Theta\,. 
\end{flalign}
The sum of the two contributions \eqref{spinning_kin} and \eqref{spinning_flux} is exactly the fermionic Lagrangian computed in Appendix D of \cite{Forini:2012bb}. 

\paragraph{The special case of the folded string.}

The folded string of \cite{Gubser:2002tv,Frolov:2002av} is obtained switching off the angle $\varphi$ in $\sphere^5$ (setting the parameter $\nu$ to zero in the classical ansatz \eqref{spinning_ansatz}). Then the  string motion is confined in $\ads_3$, {\it i.e.} $\bar t_\alpha=0$ for $\alpha=1,2$. As explained in the main body, this implies that we can choose 5 orthonormal vectors completely embedded in the compact sub-space $\sphere^5$, (in our convention \eqref{spinning_basis} they will be $\bar N_1\,, \bar N_2\,, \bar N_3\,, \bar N_4\,, \bar N_8 = N_8$ with $\nu=0$). The remaining vectors ($\hat t_\alpha$ with $\alpha=1,2$, $\hat N_1\,, \hat N_2$ and $\hat N^7=N^7$)  live only in $\ads$. Namely, the basis vectors is entirely split in the two sub-spaces. It is then immediate to see that the bosonic masses reduce to 
\be
\begin{split}
\mathcal M_{11} &=\mathcal M_{22}= - m^2_{\ads_5} = -2\,, \qquad 
\mathcal M_{77} = -2 -{2\kappa^2\omega^2\over  {\rho'}^4 }\,,\\ 
\mathcal M_{33} &=\mathcal M_{44}=\mathcal M_{55} =\mathcal M_{66}=\mathcal M_{88}= -m^2_{\sphere^5}=0\,,
\end{split}
\ee
which is the well-known results of 5 massless scalars in $\sphere^5$ and the 3 massive scalar modes living in $\ads_3 \subset \ads_5$ \cite{Frolov:2002av}. Now, the cross terms are zero $\mathcal M_{78}=0$ as well as the connection $A_\sigma^{78}=-A_\sigma^{87}=0$. 

The fermionic Lagrangian \eqref{spinning_kin} and \eqref{spinning_flux} simplifies as well. Being the normal bundle connection zero, the kinetic term reduces to 
\be
\mathcal L_{kin}= 2 i\, \bar \Theta \left( \sqrt{\gamma}\gamma^{\alpha\beta} \Gamma_a e^a_\alpha \p_\beta+ \half \rho''\Gamma_1\right)\Theta\,.
\ee
The only contribution to the flux comes from the $\ads_5$ mass in \eqref{L_ferm_flux} and the first product in \eqref{gammatilde_star_spinning2} (with $\nu=0$), thus
\be
\mathcal L_{flux} 
= i\, {\rho'}^2\, \bar\Theta\,\Gamma_{238}\Theta\,. 
\ee

\subsection{String dual to latitude Wilson loops}
\label{app:latitude}

In this Section we apply our analysis to the $\mbox{AdS}_5 \times \sphere^5$ string minimal surface corresponding to the latitude Wilson loop operators of~\cite{Drukker:2005cu, Drukker:2006ga, Drukker:2007qr}.
We choose the following patches for the target space metric:
\begin{equation}\label{metric}
\left.ds^{2}\right|_{AdS^{5}}  =-\cosh^{2}\rho\, d{t}^{2}+d\rho^{2}+{\sinh^{2}\rho\over (1+ {1\over 4} \vec{y}\cdot \vec{y})^2} \, d\vec{y} \cdot d\vec{y}\,, \qquad
\left.ds^{2}\right|_{S^{5}}  = {1\over (1+{1\over 4} \vec z\cdot \vec z)^2} d\vec z\cdot d\vec z\,, 
\end{equation}
where $\vec y=(y_1\, ,y_2\,, y_3)$ with $\vec y\cdot \vec y= y_1^2+ y_2^2+y_2^3$\,, and $\vec z= (z_1\,, \dots\,, z_5)$, with $\vec z\cdot \vec z= z_i^2+\dots +z_5^2$\,. 
The classical string solution~\cite{Drukker:2006ga} is described by the  ansatz 
\begin{equation}
\begin{split}
t &=0\,,~\rho  =\rho(\sigma)\,,~~
y^{1}  =2\sin\varphi\,,~~
y^{2}  =2\cos\varphi\,,~~
y^{3}  =0\\
z^{1} & =0\,,~~
z^{2}  =0\,,~~
z^{3}  =2\cos\theta(\sigma)\,,~~
z^{4}  =2\sin\theta(\sigma)\sin\varphi\,,~~
z^{5}  =2\sin\theta(\sigma)\cos\varphi.
\end{split}
\ee
where the worldsheet coordinates are $\varphi\in [0,2\pi)$ and $\sigma\in [0, \infty)$. The classical equations of motion read
\begin{flalign}\label{eom_latitude}
\rho' (\sigma) &= - \sinh\rho(\sigma) \,, \qquad
 \rho'' (\sigma) = \sinh\rho(\sigma)\cosh\rho(\sigma)\,, \\\nn
  \theta'(\sigma) &= \mp{1\over \cosh(\sigma_0\pm \sigma)}= \mp \sin\theta(\sigma)\,, \qquad \theta'' (\sigma)=\sin\theta(\sigma)\, \cos\theta(\sigma)\,,
\end{flalign}
with solutions given in terms of simple hyperbolic functions
\be\label{sol_hyper}
\sinh\rho(\sigma)={1\over \sinh\sigma}\,, \qquad \sin\theta(\sigma)={1\over \cosh(\sigma_0\pm \sigma)}\,.
\ee
The parameter $\s_0$ is a constant of integration defined by the internal latitude angle $\theta_0$
\be\label{constant_int_lat}
\cos\theta_0 = \tanh\s_0\,. 
\ee
To compute the quadratic fluctuation Lagrangian, we build an orthonormal basis choosing the tangent vectors  as  (from now on we only consider the solution in \eqref{eom_latitude}-\eqref{sol_hyper} with the upper sign, as it is this one that corresponds to the minimum of the action)
\begin{align}
t^A_\varphi &= \left(0,0,\cos\varphi\, \sinh \rho,-\sin\varphi\, \sinh\rho,0,0,0,0,\cos\varphi\,
   \sin\theta,-\sin \varphi\,\sin\theta\right)\,,
   \\ \nn
  t^A_\sigma &=  \left( 0,-\sinh\rho,0,0,0,0,0,\sin ^2\theta, \,-\cos \theta\,\sin\theta\,\sin\varphi,-\cos\varphi\,\sin\theta \cos\theta\right)\,,
 \end{align}
 where we omit the dependence of $\rho$ and $\theta$ on $\sigma$. The  vectors orthogonal to the worldsheet can be chosen to be 
 \begin{align}
 \bar N^A_1 & =\left(0,0,0,0,0,1,0,0,0,0\right)\,, \quad 
 \bar N^A_2 = \left(0,0,0,0,0,0,0,\cos\theta,\sin\varphi \,\sin\theta,\cos \varphi\, \sin
   \theta\right)\,,
   \\ \nn
 \bar N^A_3 & =  \left(0,0,0,0,0,0,1,0,0,0\right)\,, 
 \\ \nn
 \hat N^A_1 &= \left(-1,0,0,0,0,0,0,0,0,0\right)\,, \quad \hat N^A_2 =\left(0, 0, 0, 0, 1, 0, 0, 0, 0, 0\right)\,, \\ \nn
 \hat N^A_3 &= \left(0,0,\sin \varphi,\cos\varphi,0,0,0,0,0,0\right)\,. 
 \end{align}
The basis is completed constructing the normal vectors as
\begin{align}
N^A_7  &=
\textstyle{\frac{1}{\sqrt{\sin ^2\theta+\sinh ^2\rho}}}\left(0,0,{\cos \varphi \sin \theta },-\sin \varphi \,\sin\theta,0,0,0,0,-{\cos \varphi\, \sinh\rho},{\sin \varphi\,\sinh \rho}\right)\,,
   \\ \nn
   N^A_8 &=\textstyle{\frac{1}{\sqrt{\sin ^2\theta+\sinh ^2\rho}}}\left(0,\sin\theta,0,0,0,0,0,\sin\theta\, \sinh\rho,-\cos\theta\, \sinh\rho\, \sin\varphi
   ,-\cos\theta \sinh\rho \cos\varphi\right)\,.
\end{align}

For this  string configuration the worldsheet is space-like. The induced worldsheet metric $\gamma_{\alpha\beta}$ has then Euclidean signature and  is given by
\be
\!\!\! 
\gamma_{\alpha\beta} = \Omega^2(\s) \diag (1,  1 )\,, \qquad \Omega^2(\s) = \sinh^2\rho+\sin^2\theta= {\rho'}^2+ {\theta'}^2 = \textstyle{\frac{1}{ \cosh^2(\s+\s_0)}+\frac{1}{\sinh^2\s}}\,.
\ee
Instead, the orthonormal vectors on AdS have a time-like component, so that in this Section we will use that
\begin{align}
 \left( \hat N_i \cdot \hat N_j \right) =\eta_{ij} \,, \quad \eta_{ij}= \diag\left( -1, 1,1\right)\,, \quad i, j=1,2, 3\,,
\end{align}
while the remaining scalar products are unchanged with respect to Section \ref{sec:mass_matrix_sum_rules}. 
We report here for completeness the expressions of the worldsheet curvature $^{(2)}\!\!\, R$ and of the trace of the squared extrinsic curvature $\mathrm{Tr}(K^2)$:
\begin{equation}
^{(2)}\!\!\, R  =-\frac{2\,\partial_\sigma^2\log\Omega(\s)}{\Omega^2(\s)}
   \,, \qquad\qquad
\mathrm{Tr}(K^2) = 4\frac{ \sinh ^2(\sigma ) \cosh ^2\left(\sigma +\sigma
   _0\right)}{\cosh \left(\sigma _0\right) \cosh \left(2 \sigma
   +\sigma _0\right){}^3}\,. 
   \end{equation}

\paragraph{Bosonic Lagrangian}

We want to compute the transverse Lagrangian \eqref{L1}. We proceed first evaluating the mass matrix \eqref{wmax}. In the case of a space-like worldsheet the matrix is slightly modified: the first element in the diagonal acquires a relative sign. 
From the definition \eqref{massesads5s5} we obtain
\begin{align}\label{mdiagonal_latitude}
m^2_{\ads_5}  &= \gamma^{\alpha\beta} \left(\hat t_\alpha\cdot \hat t_\beta\right)= \frac{2 \cosh ^2(\sigma +\sigma_0)}{\cosh \sigma_0\, \cosh (2 \sigma +\sigma_0)} \,,
\\ \nn
m^2_{\sphere^5} &= - \gamma^{\alpha\beta} \left(\bar t_\alpha\cdot \bar t_\beta\right)=  -\frac{2 \sinh^2\s}{\cosh\s_0\, \cosh(2\s+\s_0)}\,\,
\end{align}
while the reduced matrix \eqref{reduced} vanishes identically. Thus, the matrix \eqref{wmax} reads
\small{
\be
\label{wmax_latitude}
\gamma^{\alpha\beta}R_{AMBN} t^{A}_{\alpha} t^{B}_{\beta} N^{i M} N^{jN}=-\begin{pmatrix} 
-m^{2}_{AdS_{5}} & 0 &0 &0&0&0&0&0\\
0 &m^{2}_{AdS_{5}} &0 &0&0&0&0&0\\
0 &0&m^{2}_{AdS_{5}}&0&0&0&0&0\\
0 &0&0&m^{2}_{S^{5}}&0&0&0&0\\
0 &0&0&0&m^{2}_{S^{5}}  &0&0&0\\
0 &0&0&0&0&m^{2}_{S^{5}} &0&0\\
0 &0&0&0&0&0&0&0\\
0 &0&0&0&0&0&0&0
\end{pmatrix}
\ee
}
\noindent
with $m^2_{\ads_5}$ and $m^2_{\sphere^5}$ given in \eqref{mdiagonal_latitude}. 

Let us now construct the projection of the extrinsic curvature along the transverse direction, {\it i.e.} $K^i_{\alpha\beta}\equiv N^i_A K^A_{\alpha\beta}$, \eqref{extrinsic}. The only non-zero components are
\begin{align}
K^7_{\varphi\sigma} &= K^7_{\sigma\varphi} =- {\sqrt{\cosh\s_0}\over \sinh\s \, \cosh(\s+\s_0)\, \sqrt{\cosh(2\s+\s_0)}} \,, \\ \nn
K^8_{\varphi\varphi}  &= K^8_{\s\s}=- {\sqrt{\cosh\s_0}\over \sinh\s \, \cosh(\s+\s_0)\, \sqrt{\cosh(2\s+\s_0)}}\,.
\end{align}
It is then straightforward to construct the whole mass matrix $\mathcal M_{ij}$ in \eqref{L1}
\begin{align}
\mathcal M_{11} &= -\mathcal M_{22}=-\mathcal M_{33}= - m^2_{\ads_5}=  \frac{2 \cosh ^2(\sigma +\sigma_0)}{\cosh \sigma_0\, \cosh (2 \sigma +\sigma_0)} \,, \\
\nn
\mathcal M_{44}&=\mathcal M_{55}=\mathcal M_{66}=m^2_{\sphere^5} = -\frac{2 \sinh^2\s}{\cosh\s_0\, \cosh(2\s+\s_0)}\,,
\\\nn
\mathcal M_{77}&=\mathcal M_{88}= {2\cosh^2(\s+\s_0)\,\sinh^2\s\over \cosh\s_0\, \cosh^3(2\s+\s_0)}\,, \quad \mathcal M_{78}=\mathcal M_{87}=0\,.
\end{align}
Finally, we calculate the normal connection $A^{ij}_\alpha$ \eqref{normalconnection}, whose non-vanishing components read
\be\label{connection_latitude}
A^{78}_\varphi = - A^{87}_\varphi= -{\tanh(2\s+\s_0)}\,.
\ee
The covariant derivatives on the transverse fields \eqref{covder} appearing in \eqref{L1} will be non-trivial only along the directions given by \eqref{connection_latitude}. 
Notice that the field-strenght $\partial_\sigma A^{78}_\varphi -\partial_\varphi A^{78}_\sigma +[A_\sigma\,,A_\varphi]^{78}= \partial_\sigma A^{78}_\varphi$ does \emph{not} vanish.



\paragraph{Fermionic Lagrangian}

The construction of the fermionic Lagrangian proceeds in two steps: the kinetic part \eqref{lagrfermkin} and the flux term \eqref{L_ferm_flux}. We perform the computations in a Lorentzian  signature for the induced worldsheet metric, and only at the end Wick-rotate back. 
For the kinetic term the only new ingredient we need is the worldsheet spin connection $\omega_{\alpha\, ab}$,
\be
\omega_{\varphi\, 12}=-\omega_{\varphi\, 21}=-\textstyle{{1\over\Omega^2(\s)}}(\textstyle{{\coth\s\over \sinh^2\s} +{\tanh(\s+\s_0)\over \cosh^2(\s+\s_0)}} )\,,
\ee
then, the final expression of the kinetic term turns out to be
\be
\tilde{L}_F^{kin}= 
2i\,\bar\Theta \, \left( \sqrt{\g}\g^{\alpha\beta}\Gamma^{a} \, e_{a\alpha}\partial_{\beta} 
-\textstyle{{1\over2\Omega(\s)}\left({\coth\s\over \sinh^2\s} +{\tanh(\s+\s_0)\over \cosh^2(\s+\s_0)}\right)}\Gamma_4+{\Omega(\s)\over 2}{\tanh(2\s+\s_0)}\Gamma_{389}\right)\,\Theta.
\ee
For the flux term \eqref{L_ferm_flux}, one obtains 
\be
\tilde{L}_F^{flux}=2i \,\bar\Theta \left( -{1\over \sinh^2\s} \Gamma_{\star}\tau_3 -{1\over \cosh^2({\s+\s_0})} \Gamma_{\star}\Gamma_{89}\right)\Theta~.
\ee
The evaluation of the determinants associated to the fluctuations above is in~\cite{FGGSVlatitude}. 

 
 \paragraph{String dual to circular Wilson loop.}
In the limit when $\theta_0\to 0$, the latitude on $\sphere^5$ shrinks to a point, and the Wilson loop reduces to a circular one, whose string dual was studied in 
\cite{Drukker:2000ep, Kruczenski:2008zk}. From equations \eqref{eom_latitude} and \eqref{constant_int_lat} the limit is equivalent to $\s_0\to\infty$ and $\theta\to \theta_0\to 0$, that is the string motion is turned off on the compact space. Taking these limits in our formulas we can reconstruct the bosonic and fermionic Lagrangian of \cite{Drukker:2000ep, Kruczenski:2008zk}. We will only change the last two normal vectors $N$, since in this way the normal connection $A$ vanishes by construction (otherwise one can always perform a rotation after to eliminate such a connection since now the corresponding field strength is flat):
\be\label{circle_n78}
N_7= \left(0\,,0\,,0\,,0\,,0\,,0\,,0\,,0\,,1\,,0\right)\,, \qquad
N_8= \left(0\,,0\,,0\,,0\,,0\,,0\,,0\,,0\,,0\,,1\right)\,.
\ee
As for the degenerate case of the folded string, here five normal vectors completely lie in $\sphere^5$ while $\ads_5$ is spanned by the two transverse vectors $\hat t_\alpha=t_\alpha$ ($\alpha=1,2$) and three normal vectors $\hat N_i$ ($i=1,2,3$). 
The rest proceeds as before, and we have
\begin{flalign} 
&^{(2)}\!\!\, R = -2\,, \qquad \mathrm{Tr}(K^2)=0\,, \qquad \Omega^2(\s)={1\over \sinh^2\s} \\ \nn
& \mathcal M_{11}= -\mathcal M_{22}= -\mathcal M_{3t3}= m_{\ads_5}^2= 2\,, \qquad
 \mathcal M_{ii}= -m^2_{\sphere^5}= 0 \,, \quad i=4\,, \dots\,, 8\,. 
\end{flalign}
  As mentioned, the choice of vectors \eqref{circle_n78} allows us to immediately eliminate the connection \eqref{connection_latitude} without resorting to any further rotation,  hence we can write
   \begin{flalign}
 \tilde{L}_F^{kin} &= 
2i \,\bar\Theta \, \left( \sqrt{\g}\g^{\alpha\beta}\Gamma^{a} \, e_{a\alpha}\partial_{\beta} 
-\textstyle{{1\over2}{\coth\s}}\Gamma_4\right)\,\Theta\,, 
\\ \nn 
\tilde{L}_F^{flux} &=-2 i\,{1\over \sinh^2\s} \,\bar\Theta\, \Gamma_{\star}\tau_3 \, \Theta~.
  \end{flalign}
  
We conclude this section by noticing that we have also carried on the computations for the minimal surface dual to generalized cusped Wilson loops \cite{Drukker:2011za}. The quadratic Lagrangian for the string fluctuations  was computed in \cite{Drukker:2011za}, and 
we have checked that also in this case we reproduces all the bosonic masses of \cite{Drukker:2011za}. 
Notice that the only non-vanishing component is the $\sigma$-component
\be 
 A^{37}_\sigma=-\frac{ \sqrt{ b^4 p^2 } \sqrt{\left(b^2+1\right) p^2-b^4}}{\sinh ^2\rho (\sigma ) \sqrt{b^4+p^2} \left(\frac{b^4-p^2}{\sinh ^2 \rho (\sigma  )}+b^2 p^2\right)}\,,
\ee
which makes this case to fall --  the solution lies both in $AdS_5$ and $S^5$ -- in the class of solutions discussed around eqs. (\ref{codazzinormal})-(\ref{reduced}). 
As $\partial_\tau  A^{37}_\sigma-\partial_\sigma A^{37}_\tau=0$, and so  the field-strength vanishes, bosonic masses are described by (\ref{reduced}) with one of the eigenvalues $\lambda_i$ vanishing.
Also, as mentioned in Section \ref{sec:fermions_kin}  a local target space rotation can be found  which eliminates the normal bundle contribution to the kinetic part of the fermionic  action (see formulas C.16-C.17 in~\cite{Drukker:2011za}).

\section*{Acknowledgments }

The work of VF and EV  is funded by DFG via the Emmy Noether Programme \emph{``Gauge Field from Strings''}.  EV acknowledges support from Research Training Group GK 1504 \emph{``Mass, Spectrum, Symmetry''} and from the Seventh Framework Programme [FP7-People-2010-IRSES] under grant agreement n. 269217 (UNIFY), and would like to thank the Perimeter Institute for Theoretical Physics for hospitality during the completion of this work. All authors would like to thank the Galileo Galilei Institute for Theoretical Physics for hospitality during the completion of this work.

\bibliographystyle{nb}
\bibliography{Ref_Fluct_Riemann}

\end{document}